\documentclass[a4paper, 10pt, usenames, dvipsnames, twoside, twocolumn]{article}

\usepackage[scaled=1.05]{newtxtext}

\usepackage[textheight={25cm},left={2.2cm},right={2.2cm},headheight=50pt,headsep=10pt]{geometry} 
\usepackage{graphicx}
\usepackage{float}
\usepackage{titlesec}
\usepackage[scaled=1.05,
 varg]{newtxmath} 

\usepackage{natbib}
\usepackage[font=small,labelfont=bf]{caption}

\usepackage{hyperref}
\usepackage[english]{babel}
\addto\extrasenglish{  
    
}
\addto\extrasenglish{  
    
}
\usepackage{mathrsfs}
\usepackage{nicefrac}
\usepackage{datetime}
\usepackage{balance}

\usepackage{amsmath}

\usepackage{setspace}
\usepackage{microtype}
\usepackage[svgnames,hyperref]{xcolor}
\hypersetup{colorlinks,linkcolor={Blue},citecolor={Blue},urlcolor={Blue}}   

\defcitealias{1917RSPSA..93..148R}{Rayleigh, 1917} 
\defcitealias{1980JAtS...37..515H}{HH}
\defcitealias{YAMAMOTOYODEN2013}{YY}

\usepackage{fancyhdr}
\titlespacing\section{0pt}{16pt}{2pt}
\titlespacing\subsection{0pt}{8pt}{2pt}
\titleformat*{\section}{\normalsize\bfseries}
\titleformat*{\subsection}{\normalsize\itshape}

\title{\vspace*{0ex}\Large\textbf{The dependence of global super-rotation on planetary rotation rate}\vspace*{.5ex}}

\author{{\large Neil T. Lewis\footnotemark[1],\, Greg J. Colyer and Peter L. Read}  \vspace*{1.25ex}\\ {\small\emph{Atmospheric, Oceanic and Planetary Physics}}  \vspace*{-.5ex}  \\ {\small\emph{Department of Physics, Clarendon Laboratory, University of Oxford, UK}} \vspace*{1.5ex} \\  {\normalsize(Draft copy, {\ddmmyydate\today})} \vspace*{-2ex}}

\date{}
\begin{document} 

\twocolumn[
  \begin{@twocolumnfalse}
    \maketitle
    \begin{abstract}
        \noindent An atmosphere may be described as globally super-rotating if its total zonal angular momentum exceeds that associated with solid-body co-rotation with the underlying planet. In this paper, we discuss the dependence of global super-rotation in terrestrial atmospheres on planetary rotation rate. This dependence is revealed through analysis of global super-rotation in idealised General Circulation Model experiments with time-independent axisymmetric forcing, compared with estimates for global super-rotation in Solar System atmospheres. Axisymmetric and three-dimensional experiments are conducted. We find that the degree of global super-rotation in the three-dimensional experiments is closely related to that of the axisymmetric experiments, with some differences in detail. A scaling theory for global super-rotation in an axisymmetric atmosphere is derived from the Held--Hou model. At high rotation rate, our numerical experiments inhabit a regime where global super-rotation scales geostrophically, and we suggest that the Earth and Mars occupy this regime. At low rotation rate, our experiments occupy a regime determined by angular momentum conservation, where global super-rotation is independent of rotation rate. Global super-rotation in our experiments saturates at a value significantly lower than that achieved in the atmospheres of Venus and Titan, which instead occupy a regime where global super-rotation scales cyclostrophically. This regime can only be accessed when eddy induced up-gradient angular momentum transport is sufficiently large, which is not the case in our idealised numerical experiments. We suggest that the `default' regime for a slowly rotating planet is the angular momentum conserving regime, characterised by mild global (and local) super-rotation. 
    \end{abstract}
    \vspace*{-1ex} {\hfill\noindent\rule{0.5\textwidth}{.75pt}\hfill}\vspace*{5ex}
  \end{@twocolumnfalse}
]{\renewcommand{\thefootnote}{\fnsymbol{footnote}}\footnotetext[1]{\href{mailto:neil.lewis@physics.ox.ac.uk}{\texttt{neil.lewis@physics.ox.ac.uk}}}}

\section{Introduction}\label{sec:intro}

Super-rotation is a phenomenon in atmospheric dynamics where the axial angular momentum of an atmosphere in some way exceeds that of the underlying planet. The magnitude of super-rotation in a planetary atmosphere may be quantified through two `super-rotation indices' \citep{1986QJRMS.112..231R,1986QJRMS.112..253R,2018AREPS..46..175R}, \begin{equation}
    S \equiv \frac{\int \rho m\,\text{d}V}{\int\rho \varOmega a^2\cos^2\vartheta\,\text{d}V}-1, \label{eq:glob_s}
\end{equation}
and
\begin{equation}
    s \equiv \frac{m}{\varOmega a^2}-1, \label{eq:loc_s}
\end{equation}
defined in terms of the axial component of specific angular momentum \begin{equation}
    m = a\cos\vartheta\left(\varOmega a\cos\vartheta +u\right). \label{eq:m}
\end{equation}
$\text{d}V=a^2\cos\vartheta\,\text{d}\lambda\,\text{d}\vartheta\,\text{d}z$ is an element of volume, and $\vartheta$, $\lambda$, and $z$ are the latitude, longitude, and geometric height coordinates, respectively. $u$ is the zonal wind velocity, and $\rho$ is the density. $a$ is the planetary radius, and $\varOmega$ is the planetary rotation rate. 

Throughout this manuscript, $S$ will be referred to as the \emph{global super-rotation index}, and $s$ the \emph{local super-rotation index}. $S$ and $s$ are defined so that $S=0$ and $s\leq0$ for an atmosphere in solid-body co-rotation with the underlying planet. It is important to note that $S$ and $s$ are distinct, and that $S$ is not simply the mass-weighted integral of $s$ over the volume of the atmosphere. The relationship between $S$ and $s$ is discussed further in \autoref{ap:gregs}. 

$S$ measures the mass-weighted global integral of $m$ relative to a state of solid-body co-rotation with the underlying planet (where $u=0$ everywhere). If $S>0$, then the atmosphere has more zonal angular momentum than if it were co-rotating with the planetary surface, and may be termed `globally' super-rotating. The requirement for $S>0$ is simply that the mass-weighted integral of $u\times a\cos\vartheta$ (that is, the relative, or `windy' contribution to $m$) is positive. 

$s$ compares $m$ at a given location with the value it would have if a parcel of air, initially at rest, was transported there from the equator whilst conserving its angular momentum. $s>0$ indicates `local' super-rotation, and at the equator is achieved when $u>0$. A maximum of angular momentum located off of the equator will typically be inertially unstable (\citetalias{1917RSPSA..93..148R}; \citealp{1981JAtS...38.2354D}), and thus the existence of local super-rotation \emph{anywhere} in an atmosphere tends to imply that there is local super-rotation at the equator (equatorial super-rotation).

In this work, we will focus on global super-rotation. Our aim is to investigate the dependence of global super-rotation on planetary rotation rate. We will explore this dependence in an idealised Earth-like general circulation model (GCM), with a similar configuration to that of \citet{1994BAMS...75.1825H}.

The rest of this work is structured as follows. In \autoref{sec:theory} we describe some constraints that may be placed on $m$, and by extension $s$ and $S$, for the case of an axisymmetric atmosphere. In \autoref{sec:previouswork} we summarise observations of super-rotation in Solar System atmospheres, and previous investigations of super-rotation conducted using numerical models. Using this material as motivation, we then proceed to conduct our investigation.  Our model configuration and experiment design is described in \autoref{sec:exp_design}. The results of our numerical simulations are presented in \autoref{sec:basic_flow} and \autoref{sec:glob_s}. Results will be presented from both three-dimensional and axisymmetric experiments. In \autoref{sec:ax3d} we explore the role of non-axisymmetric disturbances in modifying the degree of global super-rotation in the three-dimensional experiments from that of the axisymmetric experiments. In \autoref{sec:axissym} we present a scaling theory for $S$ in the axisymmetric case, obtained by appealing to the axisymmetric model of \citet{1980JAtS...37..515H}. In \autoref{sec:planets} we interpret estimates of $S$ for Solar System terrestrial atmospheres within the context of both our numerical results, and theoretical predictions for $S$. In \autoref{sec:conclude} we summarise our findings.

\section{Constraints on super-rotation strength}\label{sec:theory}

\subsection{Hide's theorem}

\begin{figure*}[!t]
    \centering\includegraphics[width=0.77\linewidth]{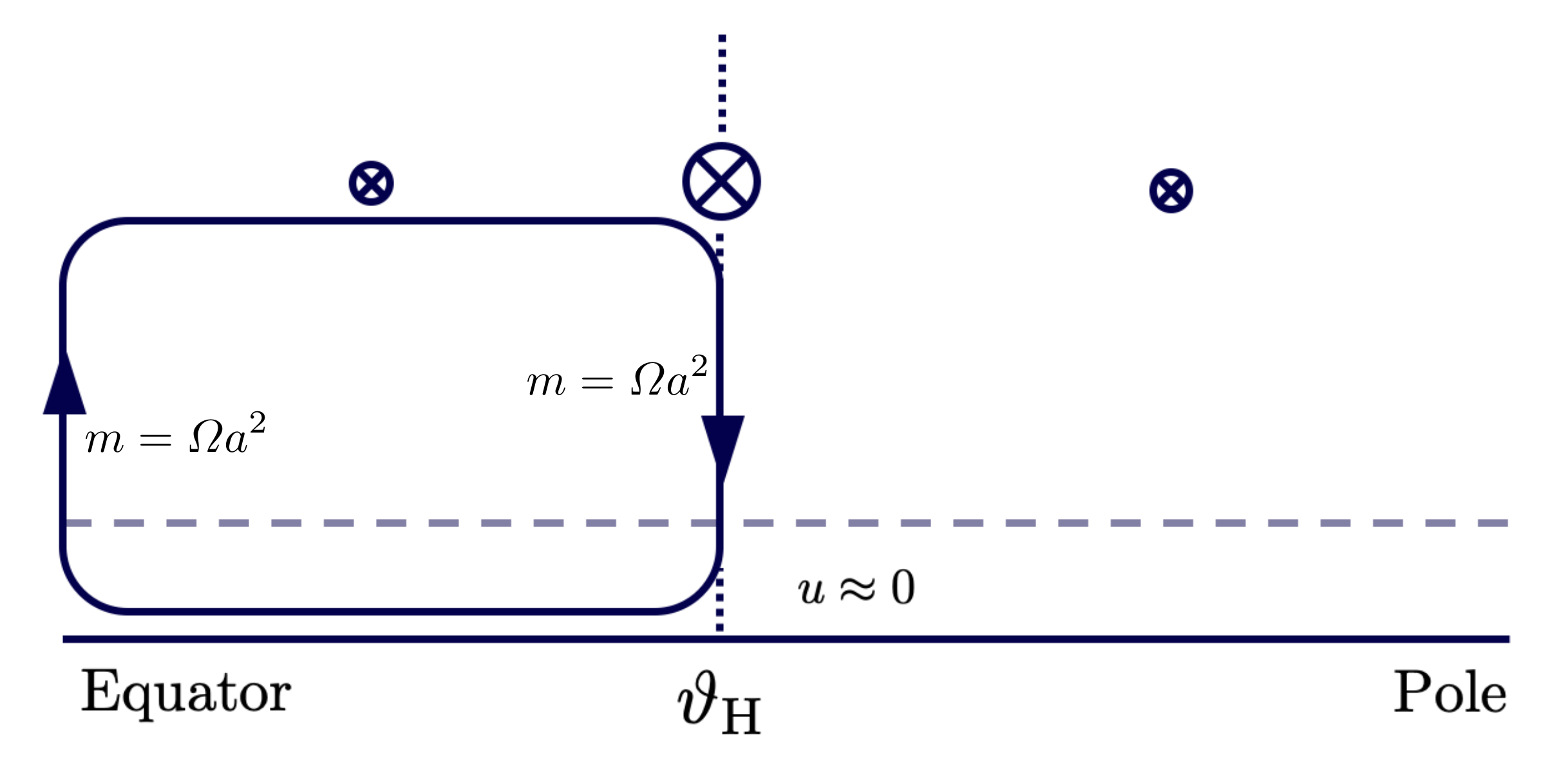}
    \caption{A schematic for the atmospheric circulation in an axisymmetric atmosphere. The atmosphere is divided into a quiescent boundary layer where $u\approx0$, and the free atmosphere where $u\neq0$. An angular momentum conserving overturning cell extends to some latitude $\vartheta_{\text{H}}$, beyond which there is no meridional overturning. The crossed circles indicate the magnitude of prograde flow aloft. }\label{fig:had_schem}
\end{figure*}

The zonally-averaged zonal momentum equation in pressure coordinates may be written \citep[][Chapter 14]{2017aofd.book.....V} \begin{align}
    \frac{\partial\overline{u}}{\partial t} +& \left(\frac{1}{a\cos\vartheta}\frac{\partial\overline{u}\cos\vartheta}{\partial\vartheta} - f\right)\overline{v}+\overline{\omega}\frac{\partial\overline{u}}{\partial p}  \label{eq:zm_u}\\ &= -\frac{1}{a\cos^{2}\vartheta}\frac{\partial}{\partial\vartheta}\left(\overline{u^{\prime}v^{\prime}}\cos^{2}\vartheta\right) - \frac{\partial \overline{u^{\prime}\omega^{\prime}}}{\partial p}\ \left(\ +\,\mathcal{V}\ \right), \nonumber 
\end{align}
where $u$ and $v$ are the zonal and meridional wind velocities, respectively, and $\omega=\text{D}p/\text{D}t$ is the pressure vertical velocity. $f=2\varOmega\sin\vartheta$ is the Coriolis parameter, and the bracketed $\mathcal{V}$ represents the possible effects of viscous friction. An overbar indicates an average over longitude (zonal average), and a prime indicates a deviation from the zonal average: \begin{equation}
    \overline{X}\equiv\frac{1}{2\pi}\int^{2\pi}_{0}X\,\text{d}\lambda\,;\qquad X^{\prime} \equiv X - \overline{X}. 
\end{equation} 
The zonally-averaged zonal momentum equation (\ref{eq:zm_u}) may be re-written in terms of $m$ \begin{equation}
    \frac{\text{D}\overline{m}}{\text{D}t}=-\frac{1}{\cos\vartheta}\frac{\partial}{\partial\vartheta}\left(\overline{m^{\prime}v^{\prime}}\cos\vartheta\right)-\frac{\partial\overline{m^{\prime}\omega^{\prime}}}{\partial p}\ \left(\,+\mathcal{V}\,\right). \label{eq:zm_m}
\end{equation}
If the atmosphere is axisymmetric ($X=\overline{X}$), and additionally inviscid, then (\ref{eq:zm_m}) reduces to \begin{equation}
    \frac{\text{D}m}{\text{D}t}=0, \label{eq:DmDt0}
\end{equation}
i.e., the specific zonal angular momentum of any given air mass is materially conserved. There is no mechanism for up-gradient transport of $m$ acting in the atmosphere, and thus it may not locally super-rotate unless it is initialised as such. This result was first presented by \citet{1969JAtS...26..841H}, and is often referred to as Hide's theorem [e.g. \citet{2017aofd.book.....V,2018AREPS..46..175R}]. %

Generalising to an axisymmetric, but viscous atmosphere (i.e. $\mathcal{V}\neq0$), \citet{1986QJRMS.112..253R} notes that \emph{molecular viscosity}, which acts towards an end-point of uniform \emph{angular velocity}, can transport angular momentum up-gradient, and so in principle could generate local super-rotation. In practise, typical molecular viscosity coefficients for gases are far too small to balance the down-gradient transport of $m$ by advection. A corollary that may be drawn here is that in the absence of an unphysically large molecular viscosity, non-axisymmetric disturbances that act to transport $m$ up-gradient are \emph{required} for the existence of a region where $s>0$, that is, a region that is locally super-rotating (assuming that no region where $s>0$ is present in the initial condition).

\subsection{Implication of Hide's theorem for global super-rotation}\label{sec:hide_S}

Consider the case of an axisymmetric atmosphere, initiated from rest. Suppose that differential heating between the equator and the pole establishes a meridional overturning (Hadley) circulation that extends to some latitude $\vartheta_{\text{H}}$, and that poleward of $\vartheta_{\text{H}}$ there is an extra-tropical region within which there is no meridional motion ($v=0$). Near the surface, friction is important and will act to restore the surface wind to a state of co-rotation with the underlying planet, so that $u\approx0$. The described circulation is sketched in \autoref{fig:had_schem}. 

In the overturning cell, parcels of air equilibrated with the surface rise at the equator, with $m=\varOmega a^{2}$. Once in the free atmosphere, (\ref{eq:DmDt0}) applies to a good approximation, $m$ is materially conserved and the Hadley cell region will be filled with air with $m=\varOmega a^{2}$. In the extra-tropical region, we do not specify the nature of the flow, but note that Hide's theorem requires $m<\varOmega a^{2}$ ($m=\varOmega a^{2}$ would imply that the overturning circulation extends beyond $\vartheta_{\text{H}}$). $s=0$ within the overturning circulation, and $s<0$ poleward of the overtuning circulation.

The fact that $s\leq0$ everywhere does not constrain $S$ to be $\leq0$. This may be verified simply by imagining the case where the overturning cell in \autoref{fig:had_schem} extends to fill the entire domain. In this scenario, $m=\varOmega a^{2}$ ($s=0$) everywhere aside from in the narrow boundary layer. Substitution of this value for $m$ into the definition for $S$ yields $S<1/2$ [if hydrostatic equilibrium is assumed; this result is easily obtained by substituting $s=0$ into \eqref{eq:a6}]. Thus, while an axisymmetric atmosphere may not permit local super-rotation, it can globally super-rotate to some degree, at least in principle.

\section{Super-rotation in Solar System atmospheres and numerical models}\label{sec:previouswork}

\subsection{Super-rotation in the Solar System}

The atmospheres of Venus and Titan (the largest moon of Saturn)  provide the clearest manifestations of both local and global super-rotation in the Solar System. Equatorial zonal wind velocities in excess of $100\,\text{m\,s}^{-1}$ have been measured in both atmospheres \emph{in situ}, by the Pioneer Venus [at Venus; \citet{1980JGR....85.8026C}], and Huygens [at Titan; \citet{2005Natur.438..800B}] descent probes, implying significant equatorial super-rotation. \citet{2018AREPS..46..175R} estimate maximum local super-rotation indices of $s_{\text{V}}\,{\sim}\,65$ and $s_{\text{T}}\,{\sim}\,15$, respectively for Venus and Titan. Estimates of the global super-rotation index for the two planets are $S_{\text{V}}\,{\sim}\,7.7$ and $S_{\text{T}}\,{\sim}\,2$ \citep{2018AREPS..46..175R}. In both cases, planetary scale equatorial disturbances have been associated with the acceleration of local super-rotation \citep{2017SSRv..212.1541S, 2014tita.book..122L}, and in the case of Venus, the semi-diurnal tide is believed to play a major role \citep{2010JGRE..115.6006L}. 

Like Venus and Titan, the atmospheres of the Earth and Mars globally super-rotate, albeit to a much lesser degree. \citet{2018AREPS..46..175R} estimate global super-rotation indices of $S_{\text{E}}\,{\sim}\,0.0135$ and $S_{\text{M}}\,{\sim}\,0.04$ for the Earth and Mars, respectively. $S>0$ may also be inferred for Jupiter and Saturn from latitudinal zonal velocity profiles [e.g., as measured by the Cassini mission; \citet{2003Sci...299.1541P,2005Sci...307.1243P}], if one assumes the profiles, which show strongly prograde jets alternating with weakly retrograde jets, to be indicative of the depth averaged weather-layer zonal velocity. For now, however, zonal velocity measurements on the two gas giant planets are too sparse in the vertical to make a quantitative estimate of their global super-rotation indices. 

Local super-rotation exists in the atmospheres of Jupiter and Saturn, where westerly equatorial jets are present with wind speeds around $60\,\text{m\,s}^{-1}$ \citep{1986Icar...65..280F,2003Sci...299.1541P} and $400\,\text{m\,s}^{-1}$ \citep{2005Sci...307.1243P,2009sfch.book..113D}, respectively, corresponding to local super-rotation indices $s_{\text{J}}\,{\sim}\,0.0075$ and $s_{\text{S}}\,{\sim}\,0.04$ \citep{2018AREPS..46..175R}. In spite of similar zonal wind velocities, local super-rotation on Jupiter and Saturn is much weaker than on Venus and Titan. Venus and Titan are smaller than Jupiter and Saturn, and rotate slower, which reduces the size of the denominator in the definition for $s$, making $s$ larger. Transient equatorial super-rotation exists in the atmospheres of the Earth and Mars, on Mars during periods of heavy dust-loading, within which the diurnal tide is amplified \citep{2003JGRE..108.5034L}, and on Earth during the westerly phase of the quasi-biennial oscillation.

\subsection{Super-rotation in numerical models with large external Rossby number}\label{sec:s_in_models}

There is now a large body of work that seeks to understand how atmospheric circulation is sensitive to key parameters and processes [as detailed in reviews by \citet{2010exop.book..471S}, \citet{2011P&SS...59..900R}, \citet{2018exop.book...PLR}, and \citet{2019jets.book....9M}]. One observation to be drawn from this work is that super-rotation may be a commonly occurring phenomenon in terrestrial\footnote[2]{When used to describe a planetary atmosphere, the term `terrestrial' refers to a shallow atmosphere above a distinct (solid or liquid) surface, where friction acts to relax the flow towards co-rotation with the underlying planet. Here, `shallow' indicates that the effective vertical extent of an atmosphere is much less than its horizontal extent. The Earth, Mars, Venus, and Titan are examples of Solar System bodies that host terrestrial atmospheres.} atmospheres. 

Numerical modelling has revealed that the external thermal Rossby number \begin{equation}
    \mathcal{R} \equiv \frac{R_{\text{d}}\Delta T_{\text{eq}}}{(\varOmega a)^{2}} \label{eq:RoT}
\end{equation}
is an important determining parameter for super-rotation strength, where strong super-rotation is associated with large $\mathcal{R}$ [reviewed in \citet{2018AREPS..46..175R}]. In (\ref{eq:RoT}), $R_{\text{d}}$ is the specific gas constant for dry air, and $\Delta T_{\text{eq}}$ is the mean radiative equilibrium equator-to-pole temperature contrast. The dependence of super-rotation strength on $\mathcal{R}$ has been investigated using idealised  three-dimensional (non-axisymmetric) GCMs, and further idealised two-dimensional `quasi-axisymmetric' models, which are zonally-averaged but include some diffusive parametrisation for eddy momentum transport which can transport angular momentum up-gradient. 

Studying a quasi-axisymmetric model of a rotating cylindrical fluid annulus, \citet{1986QJRMS.112..231R} found that $S$ scales with $\varOmega^{-2}$ at high rotation rate ($S\,{\sim}\,\mathcal{R}$) but saturates (i.e. $S=\text{const.}$) when the rotation rate is made sufficiently low. In the high rotation rate regime the zonal velocity scales geostrophically, while in the low-rotation rate regime the domain is filled with an angular momentum conserving (overturning) flow (similar to the scenario described in \autoref{sec:hide_S}). Similar behaviour has been demonstrated in a quasi-axisymmetric, Boussinesq model of the atmosphere when the horizontal eddy diffusion coefficient is not made large enough to overcome the effect of momentum advection by the overturning circulation \citep{YAMAMOTOYODEN2009}. When the strength of horizontal eddy diffusion is increased sufficiently, however, the $S=\text{const.}$ regime is replaced by one where the zonal velocity scales cyclostrophically and $S\,{\sim}\sqrt{\mathcal{R}}$ \citep{YAMAMOTOYODEN2009,YAMAMOTOYODEN2013}.

Idealised GCM experiments have shown that a terrestrial atmosphere may be expected to super-rotate at the equator `spontaneously' when $\mathcal{R}$ is made large ($\mathcal{R}\gtrsim1$). This has been achieved either by reducing the planetary radius \citep{2010JGRE..11512008M,2014JAtS...71..596P}, or rotation rate \citep{2003JAtS...60.2136W,2015ApJ...804...60K,2015JAtS...72.4281L,2018QJRMS.144.2537W,2019JAtS...76.1397C} by a significant factor. The existence of local super-rotation in these experiments implies that they are also globally super-rotating (so long as there is no strong retrograde flow elsewhere in the domain). None of these studies explicitly calculate the parameter dependence of $S$ on $\varOmega$ or $\mathcal{R}$, however.

\section{Experiment design}\label{sec:exp_design}

Our aim is to build on the work of \citet{1986QJRMS.112..231R}, \citet{YAMAMOTOYODEN2009} and \citet{YAMAMOTOYODEN2013} by studying the parameter dependence of global super-rotation on planetary rotation rate in a numerical model that solves the primitive equations on the sphere. We will present results from axisymmetric and three-dimensional model configurations. Our main goal will be to identify possible scaling regimes for $S$ in terms of $\varOmega$, and compare them with those identified in the studies listed above.  

\subsection{Numerical model}

We make use of \texttt{Isca}, a framework for building idealised general circulation models of varying complexity \citep{2018GMD....11..843V}. \texttt{Isca} is built on-top of the GFDL (Geophysical Fluid Dynamics Laboratory, Princeton) primitive equation spectral dynamical core. In the present work, \texttt{Isca} is configured as a dry-dynamical core forced by Newtonian cooling to a statically stable axisymmetric radiative-convective equilibrium temperature profile.

\subsection*{Dynamical core}

The dynamical core integrates the primitive equations forwards in time, using a semi-implicit leap-frog scheme with a Robert-Asselin time filter. The equations are solved on a spherical domain using a pseudo-spectral method in the horizontal (prognostic fields are represented by a triangular truncation of spherical harmonics), and a finite difference method in the vertical. For numerical efficiency, and to avoid formulation of vector fields on the spectral grid, the primitive equations are solved in terms of the scalar vorticity and divergence. The vertical coordinate is a terrain-following coordinate, defined as $\sigma_l = p_l/p_s$ on levels $l=1,2,\dots,80$. The levels are evenly spaced in log pressure in the troposphere, $\log \sigma\,{\propto}\,l$, and in the stratosphere the resolution is enhanced and $(\log \sigma)^{\nicefrac{1}{k}}\,{\propto}\,l$ ($k\rightarrow7.5$ as $\sigma\rightarrow0$).

\subsection*{Thermal forcing}

The relaxation temperature profile used for the Newtonian cooling is that described in \citet{2018QJRMS.144.2537W}, similar to that of \citet{1994BAMS...75.1825H}, and is written \begin{equation}
    T^{\ast} = T^{\ast}_{z}(\sigma)+T^{\ast}_{\vartheta}(\vartheta),
\end{equation}
where \begin{equation}
T^{\ast}_{z}=T^{\ast}_{z}\rvert_{\text{tp}}+\sqrt{\left[\frac{L}{2}\left(z_{\text{tp}}-z\right)\right]^{2}+K^{2}} + \frac{L}{2}\left(z_{\text{tp}}-z\right),
\end{equation}
and \begin{equation}
T^{\ast}_{\vartheta}=h(\sigma)\Delta T^{\ast}_h\left(\frac{1}{3}-\sin^{2}\vartheta\right),
\end{equation}
with \begin{equation} 
h=\begin{cases}
    \sin\left[\frac{\pi}{2}\left(\frac{\sigma-\sigma_{\text{tp}}}{1-\sigma_{\text{tp}}}\right)\right], & \text{if } \sigma \geq \sigma_{\text{tp}}, \\
    0, & \text{otherwise}.
\end{cases}
\end{equation}
$z_{\text{tp}}=12\ \text{km}$ is the tropopause height ($\sigma_{\text{tp}}$ is the corresponding $\sigma$ level), $L=6.5\ \text{K\,km}^{-1}$ is the vertical lapse rate, and $T^{\ast}_{z}\rvert_{\text{tp}}=T_0-Lz_{\text{tp}}$ is the temperature at the tropopause, with $T_0=288\ \text{K}$ the globally-averaged surface temperature. $\Delta T^{\ast}_h=60\,\text{K}$ is the equator-to-pole temperature difference at the surface. $K=2\ \text{K}$ is a smoothing parameter that ensures a continuous temperature change across the tropopause. The radiative relaxation timescale is $2.5\ \text{days}$ in the boundary layer ($\sigma>0.8$), and $30\ \text{days}$ in the free atmosphere.

\subsection*{Drag and dissipation}

A linear drag is applied in the boundary layer, operating on a timescale of $0.6$ days, and in the upper-atmosphere ($\sigma<0.01$, applied to the eddy components of the flow only), operating on a timescale of $0.5$ days. 

\begin{figure*}[!t]
    \hspace*{-4ex}\centering\includegraphics[width=1.03\linewidth]{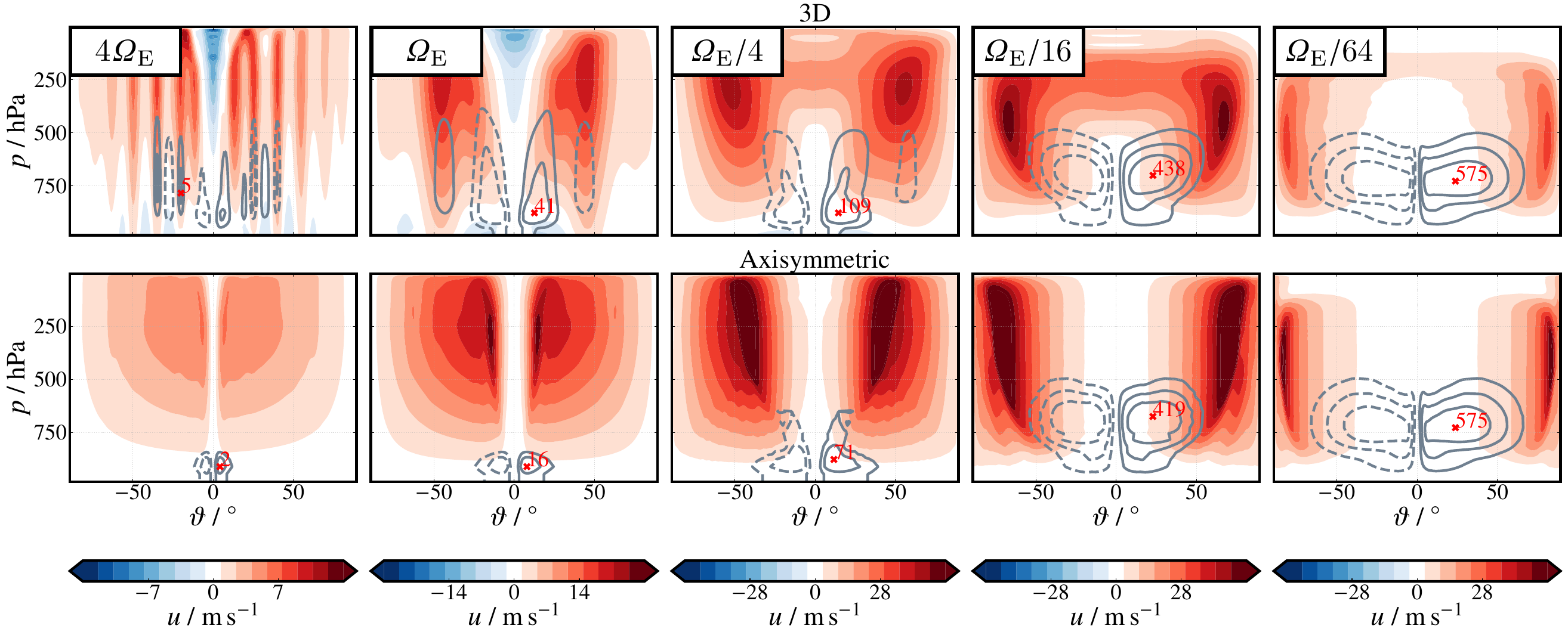}
   \caption{Zonal-mean zonal wind $u$ (colour), and meridional mass streamfunction $\Psi$ (contours). All experiments shown were run with T127 horizontal resolution. The rotation rate for each experiment is indicated in the top-left corner of each panel. Note that the colour scale for $u$ varies between panels. The red cross (and number) indicates the location (and value) where $\Psi$ is maximum. The maximum value for $\Psi$ has units $\times10^{9}\,\text{kg}\,\text{s}^{-1}$. For the $4\varOmega_{\text{E}}$, $\varOmega_{\text{E}}$, and $\varOmega_{\text{E}}/4$ experiments, the two solid contours delimit $75\%$ and $50\%$ of the maximum. For the $\varOmega_{\text{E}}/16$ and $\varOmega_{\text{E}}/64$ experiments, the $25\%$ contour is also shown. The dashed contours show `$-1\,\times$' the solid contours. As the contour levels are percentages of the maximum value for $\Psi$, they vary between panels.}\label{fig:ucomp}
\end{figure*}

A hyperviscosity term with damping order $n=4$ is applied to the horizontal momentum, and thermodynamic equations: \begin{align}
    \frac{\partial\zeta}{\partial t} &= \cdots + \nu\nabla^{2n}\zeta, \\
    \frac{\partial D}{\partial t} &= \cdots + \nu\nabla^{2n}D, \\ 
    \frac{\partial T}{\partial t} &= \cdots + \kappa\nabla^{2n}T, 
\end{align}
operating on a timescale of $0.25$ days at the grid scale. $\zeta$, $D$, and $T$ are the vorticity, divergence, and temperature. 

\subsection{Description of model runs}

We run GCM experiments over a wide range of rotation rates, from $8\varOmega_{\text{E}}$ to $\varOmega_{\text{E}}/256$, where $\varOmega_{\text{E}}=7.29\times10^{-5}\,\text{s}^{-1}$ is the Earth's rotation rate. All other planetary parameters are taken to be Earth-like: $a=6.4\times10^{6}\,\text{m}$, $g=9.81\,\text{m\,s}^{-1}$, and globally-averaged surface pressure $p_0=1\,\text{bar}$.

Three-dimensional experiments are run at both T127 and T42 spectral resolution (corresponding to approximately $1^{\circ}$ and $2.8^{\circ}$ lat-lon resolution at the equator, respectively). T127 experiments are run with rotation rates $8\varOmega_{\text{E}}$, $4\varOmega_{\text{E}}$, $2\varOmega_{\text{E}}$, $\varOmega_{\text{E}}$, $\varOmega_{\text{E}}/4$, $\varOmega_{\text{E}}/16$, and $\varOmega_{\text{E}}/64$,  and T42 experiments are run for $\varOmega=\varOmega_{\text{E}}/2^n$ for $n\in\{0,1,\dots,8\}$. Axisymmetric experiments are run for rotation rates $\varOmega=\varOmega_{\text{E}}/2^n$ for $n\in\{-2,-1,\dots,7\}$. The axisymmetric experiments are constructed by retaining only zonal wavenumber $m=0$ coefficients in the spectral dynamical core at each time step, and are run with the same meridional resolution as the T127 three-dimensional experiments.

Each of the experiments is run to equilibrium, by which we mean that $S$ is no longer evolving with time. In order to achieve this, three-dimensional experiments with $\varOmega\geq\varOmega_{\text{E}}/4$ are run for 3 (Earth) years, the $\varOmega=\varOmega_{\text{E}}/8$ experiment is run for 6 years, and experiments with $\varOmega\leq\varOmega_{\text{E}}/16$ are run for 8 years. The axisymmetric experiments are all run for 3 years.

\subsection{Data analysis}

The 3D model data analysed in this paper is output in daily-averaged format, and the axisymmetric model data is output in monthly-averaged format. \texttt{Isca} outputs data on the $\sigma$ coordinate vertical grid. We subsequently interpolate the data onto pressure levels $p_l=\sigma p_0$. Occasionally $p_l>p_s(t,\lambda,\vartheta)$, in which case the data is set to \texttt{NaN} (the grid-point being interpolated onto is below the surface). In order to correctly take account of these subterranean grid points when computing integrals and averages, we make use of the techniques described by \citet{1982MWRv..110.1801B}.

\section{Zonally-averaged circulations}\label{sec:basic_flow}

To provide context for the rest of the paper, we begin the presentation of our experiment results by discussing some aspects of their zonally-averaged circulations. 

Each of the experiments discussed in this section was run at T127 resolution. 
\subsection{Zonal wind and meridional overturning}

The zonal-mean zonal wind $\overline{u}$, and meridional mass streamfunction \begin{equation}
    \Psi\left(\vartheta,\,p\right) = 2\pi a\cos\vartheta\int^{0}_{p}\overline{v}\,\text{d}p/g \label{eq:psi}
\end{equation}
are shown in \autoref{fig:ucomp} for the 3D (top row) and axisymmetric (bottom row) experiments, over a range of rotation rates. The zonal-mean zonal wind in the 3D $\varOmega=\varOmega_{\text{E}}$ experiment consists of a sub-tropical jet, associated with angular momentum conservation in the Hadley cell, and an `eddy-driven' jet, accelerated by Rossby waves generated by mid-latitude baroclinic instability, which converge momentum into their source region \citep{1980JAtS...37.1216T}. As the rotation rate is increased to $4\varOmega_{\text{E}}$, the number of eddy driven jets increases, as the characteristic lengthscale of baroclinic instability decreases \citep{2005JAtS...62.2484L}. At rotation rates $\leq\varOmega_{\text{E}}/4$, a prograde (super-rotating) jet emerges at the equator. This is associated with planetary-scale tropical disturbances which converge momentum towards the equator \citep{2010JGRE..11512008M}, which are generated when $\mathcal{R}$ is increased to an $\mathcal{O}(1)$ value \citep{2014GeoRL..41.4118W}.

Axisymmetric theory \citep{1980JAtS...37..515H} predicts that the meridional extent of the Hadley circulation scales inversely with the planetary rotation rate, at high and moderate (similar to the Earth) rotation rates, before saturating as the Hadley cell begins to fill the domain at low rotation rate \citep{1984JAtS...41.3437H}. Concomitantly, as $\varOmega$ is reduced the maximum zonal velocity of the sub-tropical jet (produced by conservation of angular momentum within the Hadley cell) at first increases, as the jet moves polewards, before decreasing with $\varOmega$ in the low rotation rate limit, when the jet has essentially been pushed as far polewards as possible \citep{2019JAtS...76.1397C}. Both of these effects may be seen in the results from our axisymmetric experiments. 

In the non-axisymmetric (3D) experiments, the Hadley cell width also increases as $\varOmega$ is decreased, before saturating at low rotation rates. This is partially due to the underlying axisymmetric processes described in the previous paragraph, and also because the meridional extent of the Hadley circulation is limited by baroclinic instability \citep{1967nthr.book.....L}. The latitude at which the angular momentum conserving flow in the Hadley circulation becomes unstable to baroclinic instability scales inversely with $\varOmega$ to some power \citep{2006JAtS...63.3333W}. [Specifically, using a two-layer quasi-geostropic model \citet[][Chapter 14]{2017aofd.book.....V} derives a $\varOmega^{-\nicefrac{1}{2}}$ scaling for the Hadley cell latitude (in the small-angle approximation, on the $\beta$-plane).] 

As in the axisymmetric experiments, the sub-tropical jet velocity in the 3D experiments increases up until $\varOmega=\varOmega_{\text{E}}/16$, before decreasing. At all rotation rates shown (with the exception of $4\varOmega_{\text{E}}$), the maximum  sub-tropical jet velocity in the 3D experiments is less than that in the axisymmetric experiments. At low rotation rates, the sub-tropical jets are decelerated by the equatorial disturbances which converge momentum to the equator, and at high rotation rates, the Rossby waves that accelerate the eddy-driven jets dissipate on the poleward flanks of the sub-tropical jets and decelerate the flow there.

We note that the rotation rate dependence of $\overline{u}$ and $\Psi$ described in this section is broadly consistent with that presented in previous works [e.g., \citet{2015ApJ...804...60K}, \citet{2018QJRMS.144.2537W}, and \citet{2019JAtS...76.1397C}].

\subsection{Local super-rotation index}

\begin{figure*}[!t]
    \hspace*{-4ex}\centering\includegraphics[width=1.03\linewidth]{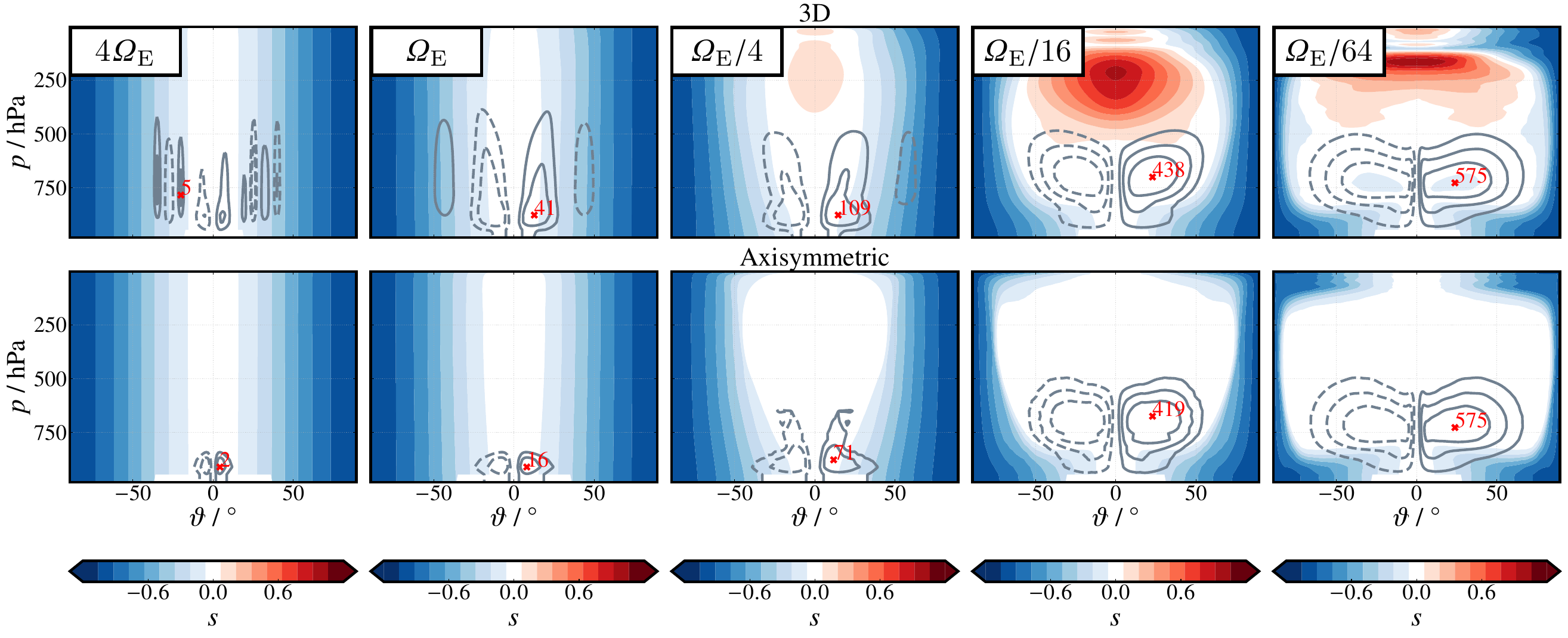}
    \caption{Local super-rotation index, $s$. As in \autoref{fig:ucomp}, the red label and contours show the meridional mass streamfunction $\Psi$ (see \autoref{fig:ucomp} caption for details). All experiments shown were run with T127 horizontal resolution.}\label{fig:loc_s}
\end{figure*}

\autoref{fig:loc_s} presents the local super-rotation index $s$ for the same set of experiments shown in \autoref{fig:ucomp}, and $\Psi$ is once again overlaid for reference. 

We begin by briefly commenting on the axisymmetric experiments. The action of the Hadley cell to fill the atmosphere with air equilibrated with the equatorial surface (e.g. as illustrated in \autoref{fig:had_schem}) is made clearly visible by the white region where $s\approx0$ in each of the panels. This region grows in meridional extent as the Hadley cell expands when $\varOmega$ is reduced. Local super-rotation ($s>0$) is not present in any of the axisymmetric experiments, in accordance with Hide's theorem.

In the rapidly rotating 3D experiments ($\varOmega\geq\varOmega_{\text{E}}$) local super-rotation is also absent. This is because although non-axisymmetric disturbances are present in the atmosphere (as evidenced by the presence of eddy driven jets in the $\overline{u}$ wind fields), they act to transport angular momentum down its local gradient. At high rotation rate, $s$ in the 3D experiments is very similar to that in the axisymmetric experiments. This is largely because the `windy' ($u\times a\cos\vartheta$) contribution to $m$ from the eddy-driven jets is small with respect to the contribution to $m$ from the background rotation of the planet ($\varOmega a^{2}\cos^{2}\vartheta$).

Once the rotation rate is reduced to $\varOmega_{\text{E}}/4$, local super-rotation maximal at the equator emerges. This indicates that non-axisymmetric disturbances which induce up-gradient transport of $m$ are now present. It is of interest to note that whilst sub-tropical and equatorial zonal velocities in the $\varOmega_{\text{E}}/4$ and $\varOmega_{\text{E}}/16$ experiments are similar (\autoref{fig:ucomp}), the maximum local super-rotation \emph{relative} to the underlying planet is much weaker in the $\varOmega_{\text{E}}/4$ experiment, when compared with the $\varOmega_{\text{E}}/16$ experiment. The equatorial zonal wind velocity in the $\varOmega_{\text{E}}/16$ experiment appears much greater than that in the $\varOmega_{\text{E}}/64$ experiment, however the maximum local super-rotation is similar in the two experiments.

In each of the simulations where local super-rotation is present, $s$ is only significantly positive in the region above the Hadley circulation. This is because within the Hadley cell, air ascending at the equator acts to communicate the effects of friction at the surface to the atmosphere aloft (i.e. it is advecting air with $s=0$). This decelerates any zonal flow that forms on or near the equator within the Hadley Cell.

\section{Dependence of global super-rotation on rotation rate}\label{sec:glob_s}

\begin{figure*}[!t]
    \centering\includegraphics[width=.90\linewidth]{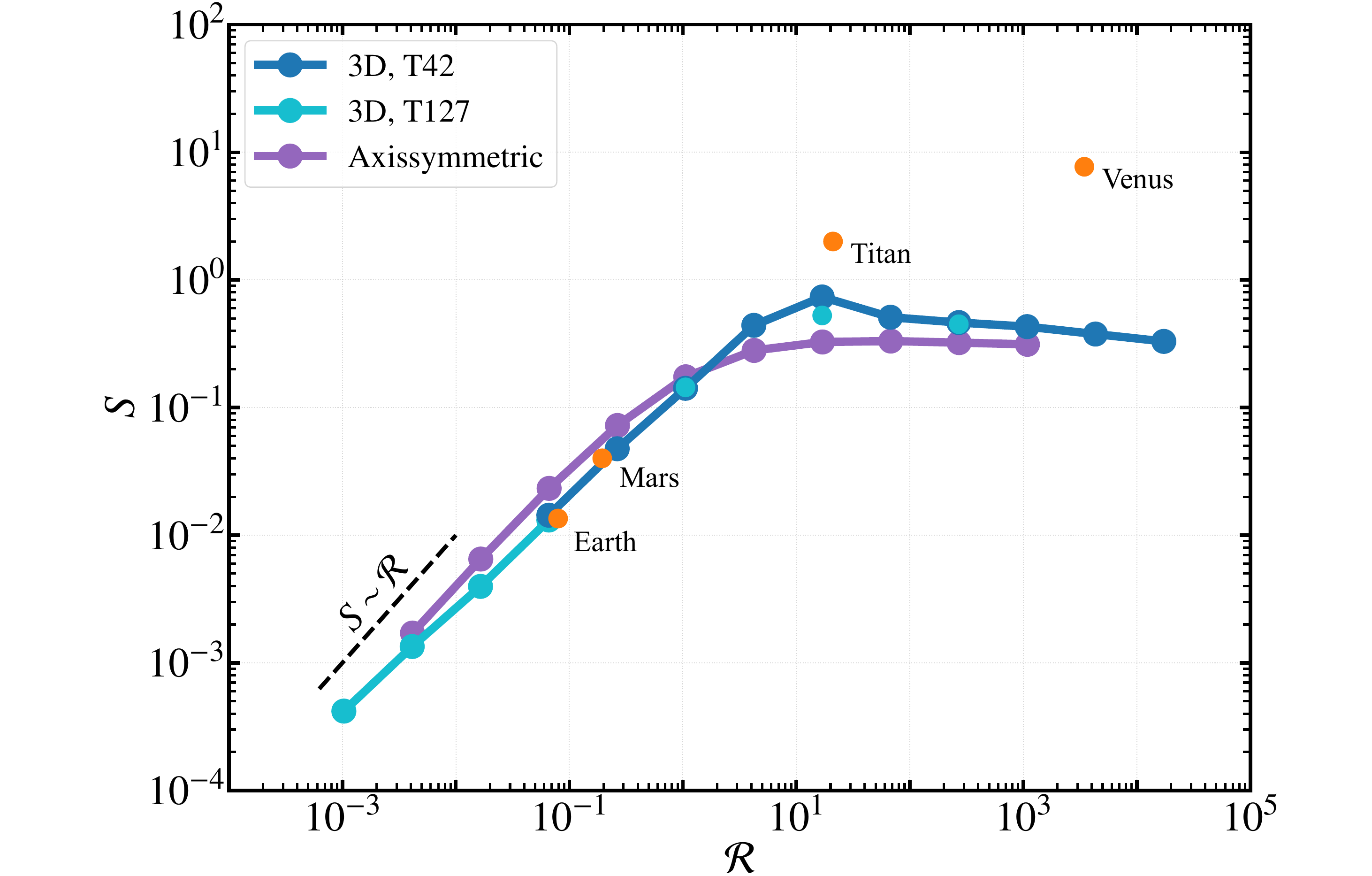}
    \caption{Global super-rotation index $S$ vs. thermal Rossby number $\mathcal{R}$. 3D experiments are shown in blue, and axisymmetric experiments in purple. Experiments run at T42 are plotted in dark blue, and those run at T127 are plotted in light blue. The orange dots show estimates for $S$ for the Earth, Mars, Titan and Venus, and are taken from \citet{2018AREPS..46..175R}. The black dashed line indicates a $\varOmega^{-2}$ scaling.}\label{fig:the_graph}
\end{figure*}

We now turn our discussion towards the dependence of the global super-rotation index $S$ on planetary rotation rate $\varOmega$. \autoref{fig:the_graph} shows $S$ vs. $\mathcal{R}$ for each of the experiments run for this work. We choose to present $S$ as a function of $\mathcal{R}$ instead of $\varOmega$ in order to make a cleaner comparison between our numerical experiments, and estimates of $S$ for the real planets (Venus, Titan, Mars, the Earth). This is necessary as the real planets are of different sizes, in addition to having different rotation rates, and these parameters can be of similar dynamical importance [see \citet{2010JGRE..11512008M} and \citet{2014Icar..238...93D}]. The values used to estimate $\mathcal{R}$ for our numerical experiments and the real planets are provided in \autoref{tab:RoT}.

The axisymmetric and 3D experiments reveal a similar dependency for $S$ on $\mathcal{R}$. At low $\mathcal{R}$ (high rotation rate), $S$ increases with decreasing $\varOmega$. Global super-rotation in the axisymmetric experiments ($S_{\text{ax}}$) appears to scale inversely with $\varOmega^{2}$; this may be inferred by comparing the black dashed line ($\varOmega^{-2}$) with the solid purple line ($S_{\text{ax}}$). At high $\mathcal{R}$ (low rotation rate), $S_{\text{ax}}$ appears to saturate around a value $S\approx0.3\,{-}\,0.5$. These high- and low-rotation rate scalings appear to correspond with those identified by \citet{1986QJRMS.112..231R} in a quasi-axisymmetric model of a cylindrical fluid annulus, and \citet{YAMAMOTOYODEN2009} in a quasi-axisymmetric Boussinesq model on a sphere (when up-gradient diffusive transport of angular momentum is small).

The values taken by $S$ in the 3D experiments ($S_{3\text{D}}$) appear to be perturbations about $S_{\text{ax}}$.  In the rapidly rotating regime, $S_{3\text{D}}$ is lower than $S_{\text{ax}}$, while in the slowly rotating regime, $S_{3\text{D}}$ is greater than $S_{\text{ax}}$. The excess global super-rotation achieved by the 3D experiments in the slowly rotating regime is indicative of the local super-rotation in these experiments, which could be interpreted as being superimposed on the underlying axisymmetric circulation (\autoref{fig:loc_s}). 

\begin{table}[!b]
    \small 
    \def\arraystretch{1.5}
    \centering\begin{tabular}{|l|cccc|c|}
        \hline
        Planet & $a$ & $\varOmega$ & $\Delta T_{\text{eq}}$ & $R_{\text{d}}$  & $\mathcal{R}$ \\ 
        \hline
        Earth & $6400$ & $7.29\times10^{-5}$  & $60$ & 287 & 0.07 \\ 
        \hline 
        Mars  & $3396$ & $7.09\times10^{-5}$ & $60$ & 188 & 0.20 \\
        \hline 
        Venus & $6051$ & $2.99\times10^{-7}$  & $60$ & 188 & 574 \\  
        \hline 
        Titan & $2575$ & $4.56\times10^{-6}$  & $10$ & 290 & 21 \\ 
        \hline 
        \texttt{Isca} & $6400$ & variable & $60$ & 287 & variable \\ 
        \hline 
    \end{tabular}
    \caption{Parameter values used to estimate $\mathcal{R}$ for \autoref{fig:the_graph}.  \\ Units: km for $a$, $\text{s}^{-1}$ for $\varOmega$ , $\text{K}$ for $\Delta T_{\text{eq}}$, $\text{J\,kg}^{-1}\text{\,K}^{-1}$ for $R_{\text{d}}$.}\label{tab:RoT}
\end{table}

Estimates of $S$ for the Earth, Mars, Venus, and Titan [taken from \citet{2018AREPS..46..175R}] are shown as orange dots in \autoref{fig:the_graph}. The 3D experiments with $\mathcal{R}$ closest to the Earth ($\varOmega_{\text{E}}$) and Mars ($\varOmega_{\text{E}}/2$) obtain values for $S$ similar to those obtained by the two real planets. At low rotation rate, however, experiments with $\mathcal{R}$ characteristic of Venus ($\varOmega_{\text{E}}/128$) and Titan ($\varOmega_{\text{E}}/16$) fail to generate global super-rotation of a similar order of magnitude to that estimated for the real planets.

The dependence of $S$ on $\mathcal{R}$ presented in \autoref{fig:the_graph} raises a number of questions. What processes contribute to the scaling for $S$ that appears in the rapidly and slowly regimes? Why is $S_{3\text{D}}<S_{\text{ax}}$ at high rotation rates, and $S_{3\text{D}}>S_{\text{ax}}$ at low rotation rates? Finally, why does $S$ saturate for $\mathcal{R}\gtrsim1$ in our numerical experiments, and how does the low rotation rate behaviour in our three-dimensional experiments relate to the atmospheric circulations of Venus and Titan? In the following three sections of this paper, we seek to tackle these questions. In \autoref{sec:ax3d} we will investigate how non-axisymmetric disturbances in the 3D experiments modify the zonal angular momentum budget from the axisymmetric case, leading to the modification of $S$. In \autoref{sec:axissym} we derive a scaling for $S$ in the axisymmetric case from the \citet{1980JAtS...37..515H} model. In \autoref{sec:planets} we discuss the estimates of global super-rotation for the Solar System atmospheres in the context of our idealised numerical experiments, axisymmetric theory, and an alternative theory proposed by \citet{YAMAMOTOYODEN2013} for atmospheres with large up-gradient transport of angular momentum.

\section{Eddy effects on global super-rotation}\label{sec:ax3d}

In this section, we would like to understand the cause of differences in the value of $S$ achieved by the 3D and axisymmetric experiments. To do so, we will appeal to the Gierasch--Rossow--Williams mechanism [after \citet{1975JAtS...32.1038G} and \citet{1979JAtS...36..377R}]. Specifically, we would like to understand why $S_{3\text{D}}>S_{\text{ax}}$ in the slowly rotating regime. Quantitative analysis in this section will make use of the T127 $\varOmega_{\text{E}}/16$ experiment. A comment about the rapidly rotating regime where $S_{\text{ax}}>S_{3\text{D}}$ will be made at the end of the section. 

\subsection{Description of the model spin-up phase}

Each of the numerical experiments studied in this work is initiated at rest, that is with $u=0$ everywhere, corresponding to $S=0$. The value for $S$ obtained by each experiment in equilibrium is a measure of the total zonal angular momentum injected into the atmosphere during spin-up. 

\begin{figure}[!b]
    \hspace{-3ex}\centering\includegraphics[width=\linewidth]{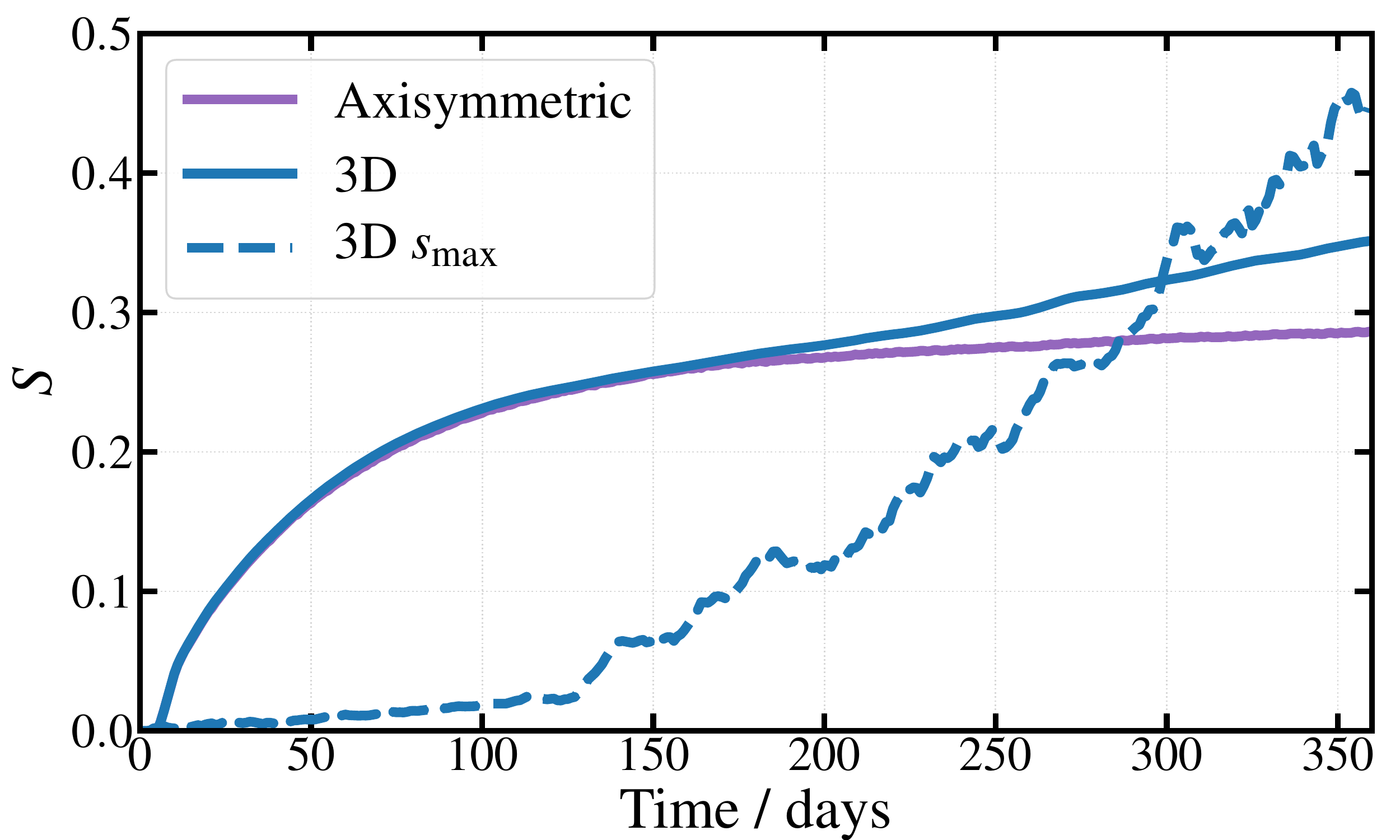}
    \caption{$S$ vs. $t$ for the first 360 days of the $\varOmega_{\text{E}}/16$ experiments. The blue curve shows the 3D experiment and the purple curve shows the axisymmetric experiment. $\max(s)$ is shown for the 3D experiment as a dashed curve.}\label{fig:dS_dt}
\end{figure}

\begin{figure}[!t]
    \hspace*{-3ex}\centering\includegraphics[width=\linewidth]{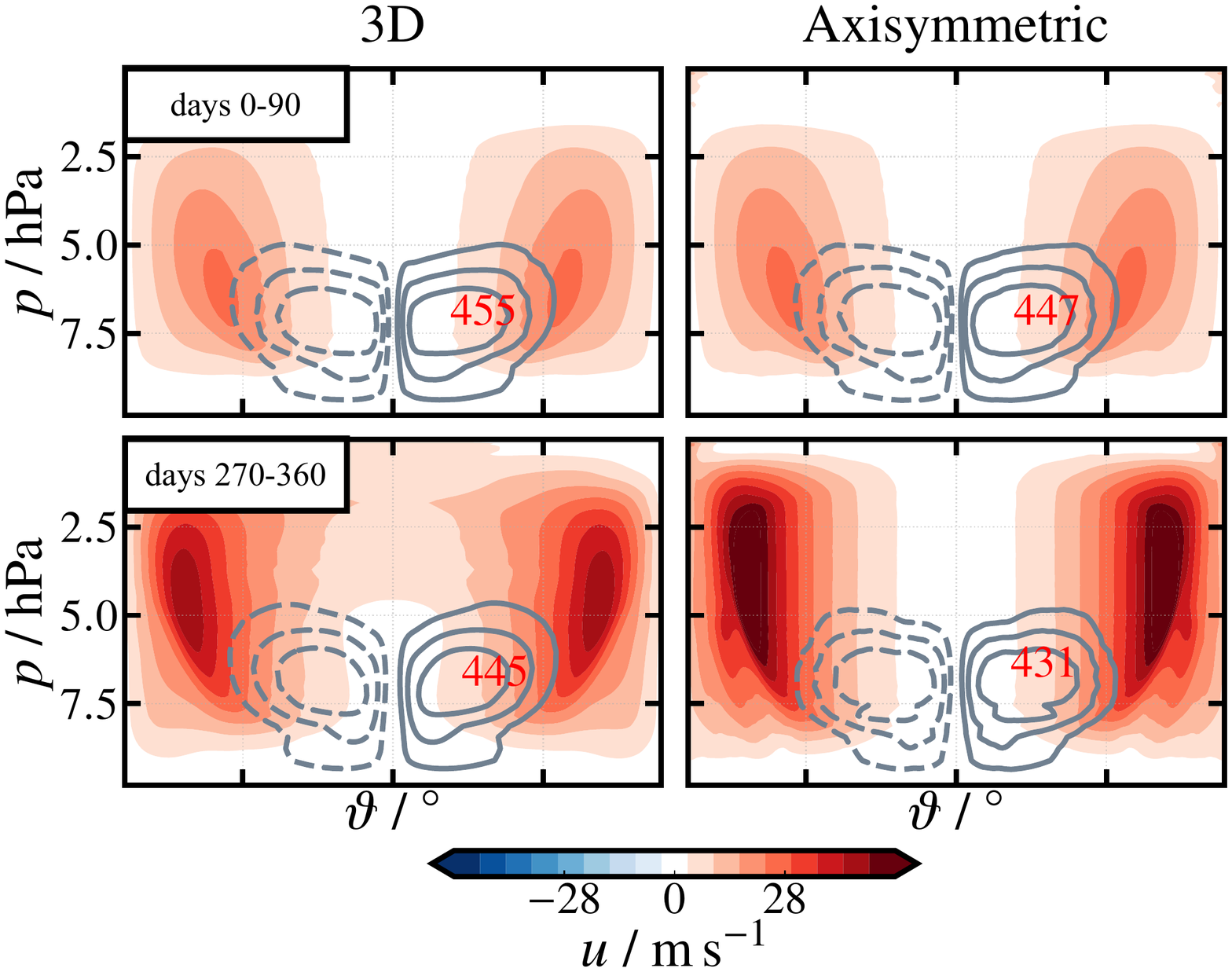}
    \caption{$u$ and $\Psi$ shown for the 3D and axisymmetric $\varOmega_{\text{E}}/16$ experiments during spin-up. The top row is averaged over days 0-90, the bottom row is averaged over days 270-360. The meaning of the contours for $\Psi$ is as in \autoref{fig:ucomp}. The data shown is from experiments run with T127 horizontal resolution.}\label{fig:ax_vs_3D_zm}
\end{figure}

\autoref{fig:dS_dt} shows $S$ as a function of time for the first 360 days of the 3D and axisymmetric $\varOmega_{\text{E}}/16$ experiments. For the first 150 days of model run time, $S$ in the 3D and axisymmetric cases is essentially the same. During this period, the 3D experiment goes through an `axisymmetric phase' during which the zonally-averaged circulations in the 3D and axisymmetric experiments are indistinguishable. This is shown in \autoref{fig:ax_vs_3D_zm} which presents $u$ and $\Psi$ averaged over days 0-90 and 270-360 for the 2D and 3D $\varOmega_{\text{E}}/16$ experiments. 

After the axisymmetric phase, non-axisymmetric disturbances emerge in the 3D experiment which act to flux angular momentum up its local gradient. This leads to the emergence of local super-rotation (see \autoref{fig:ax_vs_3D_zm}, days 270-360), and correspondingly $S_{3\text{D}}$ and $S_{\text{ax}}$ diverge. The maximum local super-rotation index in the 3D experiment is shown as a blue dashed line in \autoref{fig:dS_dt}.

\begin{figure}
    \centering\includegraphics[width=.77\linewidth]{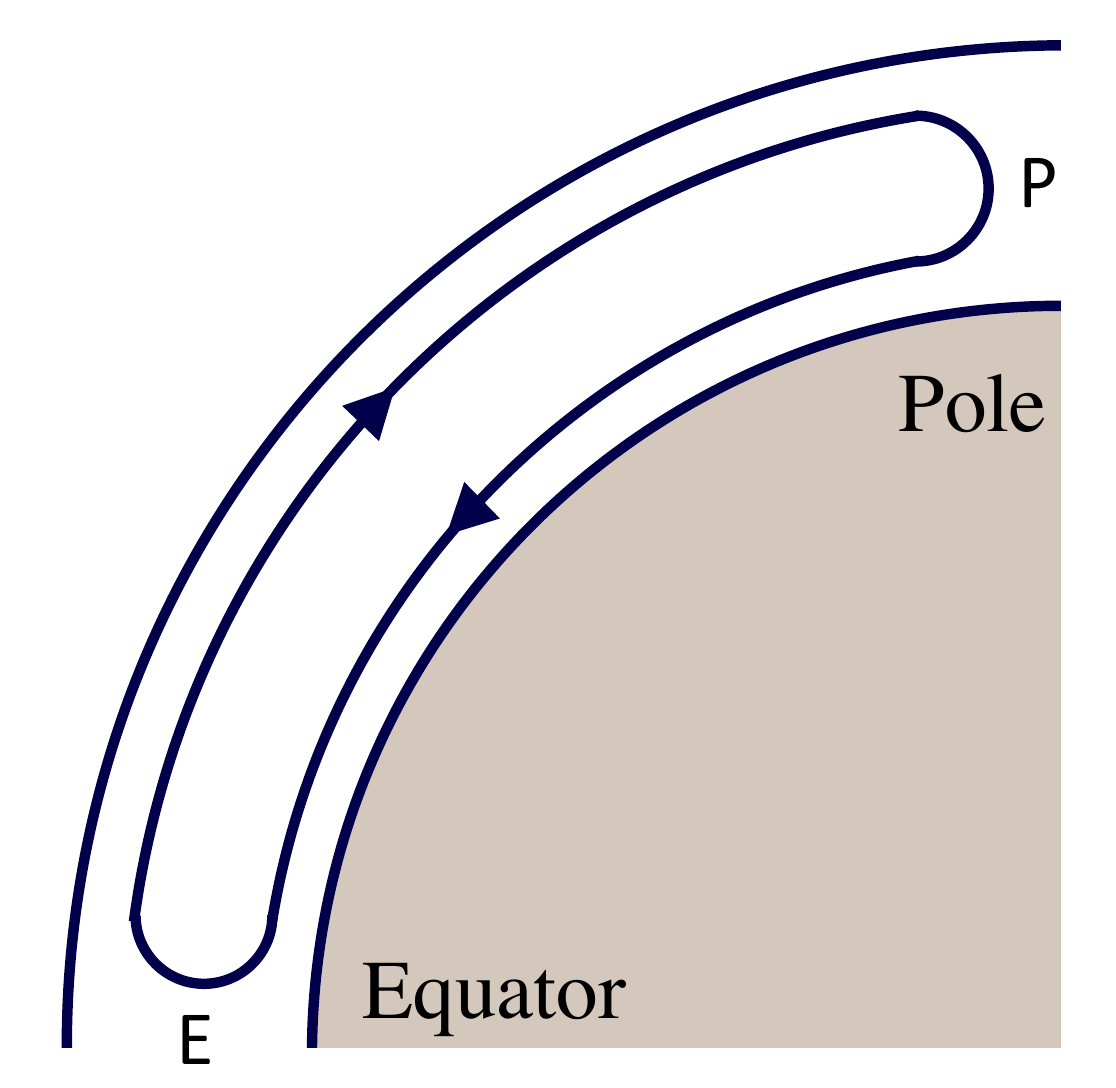}
    \caption{Schematic of a meridional cell, after \citet{1975JAtS...32.1038G}. If the upward flux of $m$, \textsf{E} is greater than the downward flux, \textsf{P}, the cell produces a net upward vertical flux of $m$.}\label{fig:gierasch}
\end{figure}

\subsection{Net upward transport of angular momentum by the Hadley circulation}

\begin{figure*}[!ht]
    \centering\includegraphics[width=.82\linewidth]{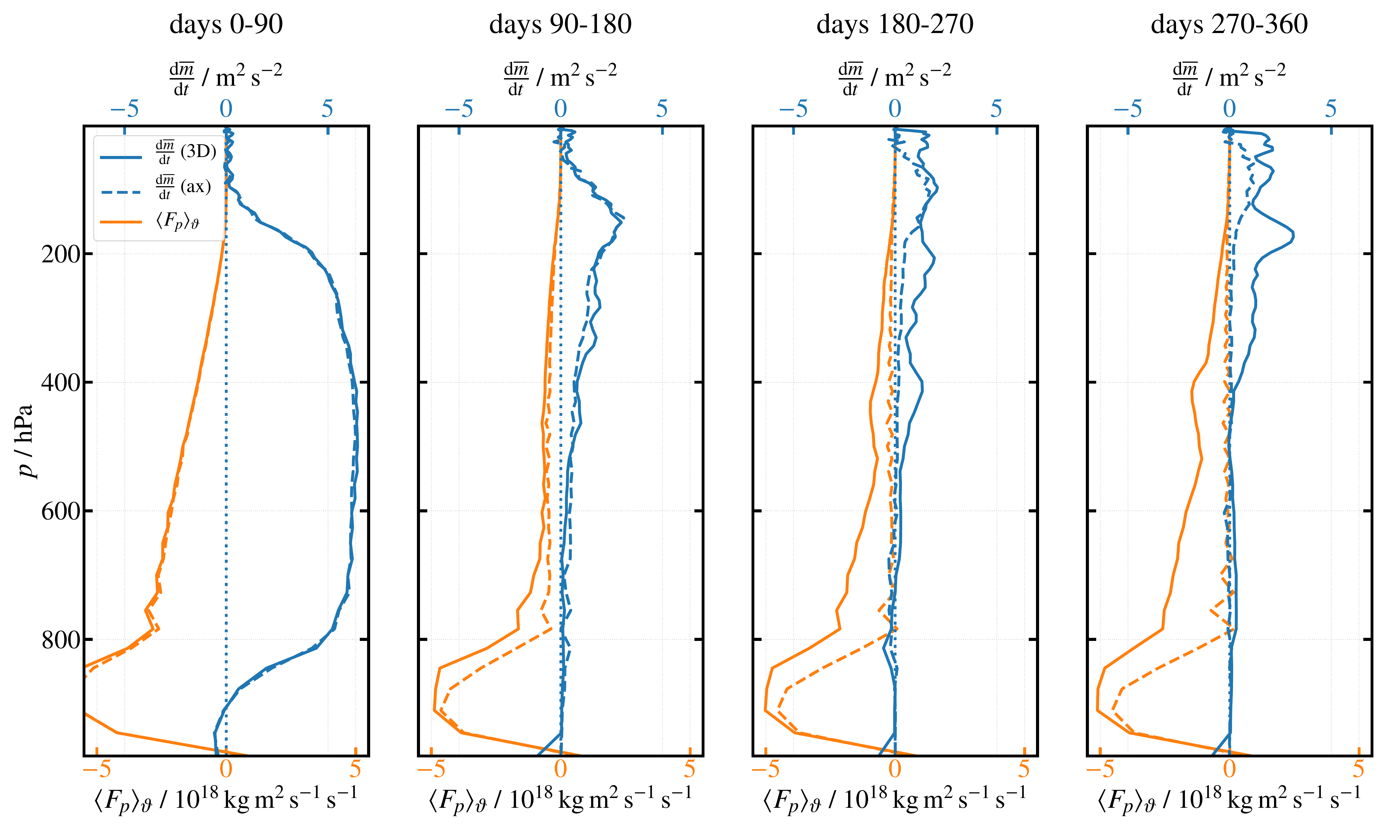}
    \caption{Meridionally averaged zonal mean zonal angular momentum tendency $\text{d}\overline{m}/\text{d}t$ (blue) and total vertical momentum flux by the residual circulation $\langle F_p\rangle_{\vartheta}=g^{-1}\int\overline{\omega}^{\ast}\overline{m}\,a^{2}\cos\vartheta\,\text{d}\vartheta$ (orange) for the $\varOmega_{\text{E}}/16$ experiments. Solid lines are for the 3D experiment, dashed lines are for the axisymmetric experiment.}\label{fig:w_mom_flux}
\end{figure*}

When the numerical experiments are begun, heating over the equator and cooling in polar regions leads to the establishment of a Hadley circulation. Air, initially at rest, moves towards the equator in the surface branch; if this air conserves its angular momentum, then it will acquire negative zonal velocity. This air with $u<0$ then experiences frictional drag, which acts to return the air towards $u=0$. As the equatorward moving flow initially has $u<0$, this corresponds to an injection of positive $u$, or $m$ into the atmosphere. Air with $m=\varOmega a^{2}$ (the equatorial value for $m$ when $u=0$) is then transported upward and poleward by the angular momentum conserving overturning circulation until the entire free atmosphere is filled with air with $m=\varOmega a^{2}$. 

In the axisymmetric case, once an atmosphere with $m=\varOmega a^{2}$ everywhere is established, $S$ can no longer increase. This is because there is no longer an exchange of angular momentum across the boundary between the free atmosphere and the boundary layer (e.g. air leaving and entering the boundary layer has $m=\varOmega a^{2}$). 

In the 3D experiment, however, the emergence of non-axisymmetric disturbances which flux momentum equatorward can modify the angular momentum budget within the Hadley circulation, so as to allow it to further extract angular momentum from the planetary boundary layer.

Analysis of wave-mean flow interaction is most clearly performed using the transformed Eulerian mean (TEM) formulation of the zonal mean momentum equation \citep{1976JAtS...33.2031A, 1978JAtS...35..175A}. Within this formulation, the zonal mean zonal angular momentum equation (\ref{eq:zm_m}) is re-written \begin{equation}
    \frac{\partial\overline{m}}{\partial t} + \overline{\pmb{v}}^{\ast}\cdot\nabla\overline{m}=\nabla\cdot\pmb{F}, \label{eq:TEM}
\end{equation}
where 
\begin{align}
    \pmb{F}&=\left\{ F^{\prime}_\vartheta,\  F^{\prime}_p\right\} \nonumber \\ 
           &=\left\{ -\overline{m^{\prime}v^{\prime}}+\psi\frac{\partial\overline{m}}{\partial p},\ \   -\overline{m^{\prime}\omega^{\prime}}-\frac{\psi}{a}\frac{\partial\overline{m}}{\partial\vartheta}\right\} \label{eq:ep_flux}
\end{align}
is the Eliassen-Palm flux [EP flux, after \citet{1961GeoPl...5...19E}], with \begin{equation}
    \psi = \left.\overline{v^{\prime}\theta^{\prime}}\middle/\frac{\partial\overline{\theta}}{\partial p}\right.,
\end{equation}
and $\overline{\pmb{v}}^{\ast}=(0,\overline{v}^{\ast},\overline{\omega}^{\ast})$ is a residual circulation, defined by 
\begin{equation}
    \overline{v}^{\ast} \equiv \overline{v} - \frac{\partial\psi}{\partial p},\quad\overline{\omega}^{\ast}\equiv\overline{\omega}+\frac{1}{a\cos\vartheta}\frac{\partial\left(\psi\cos\vartheta\right)}{\partial\vartheta}.\label{eq:res_circ}
\end{equation}
The residual circulation $\pmb{v}^{\ast}$ describes the component of the conventional Eulerian mean circulation whose contribution to temperature change is not cancelled by eddy heat flux divergence \citep[][Chapter 10]{holton1973introduction}. The residual circulation approximates that driven by \emph{diabatic} heating, and thus more closely resembles the Lagrangian mean meridional mass flow than the conventional Eulerian mean meridional circulation \citep[][both Chapter 10]{holton1973introduction,2017aofd.book.....V}. The EP flux divergence $\nabla\cdot\pmb{F}$ describes the total wave induced force on the zonally-averaged zonal flow. 

\begin{figure}[!ht]
    \centering\includegraphics[width=.86\linewidth]{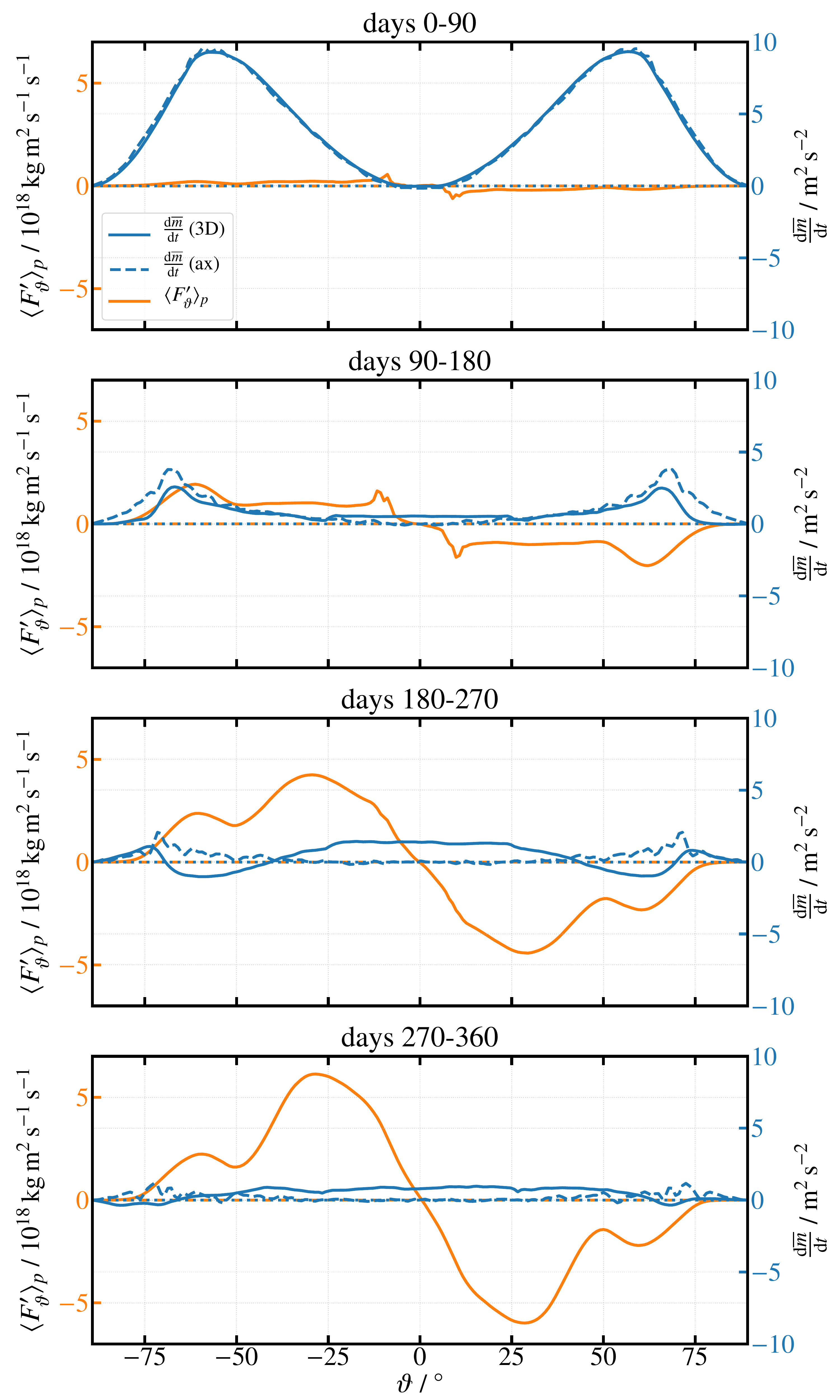}
    \caption{Vertically averaged $\text{d}\overline{m}/\text{d}t$ (blue) and total merdional momentum flux due to wave induced forces $\langle F_\vartheta^{\prime}\rangle_{p}=2\pi a\cos\vartheta\int F^{\prime}_{\vartheta}\,\text{d}p/g$ (orange) for the $\varOmega_{\text{E}}/16$ experiments. Solid lines are for the 3D experiment, dashed lines are for the axisymmetric experiment.}\label{fig:v_mom_flux}
\end{figure}

\autoref{fig:gierasch} presents a schematic of a meridional overturning cell; upward transport of $m$ in the ascending branch of the circulation is symbolised by \textsf{E}, and downward transport in the descending branch by \textsf{P}. \autoref{fig:w_mom_flux} and \autoref{fig:v_mom_flux} show plots of the total vertical momentum flux by the mean flow\footnote[3]{This expression may be obtained by noting that the TEM zonal angular momentum budget (\ref{eq:TEM}) may be written \begin{equation}
    \frac{\partial\overline{m}}{\partial t} + \nabla\cdot\left(\overline{\pmb{v}}^{\ast}\overline{m}\right)=\nabla\cdot\pmb{F} \nonumber 
\end{equation}
by making use of the fact that the residual circulation is divergence free. Acceleration of the zonal mean flow in the free atmosphere is then due to a balance between convergence and divergence of fluxes of $\overline{m}$ by the residual mean circulation ($\overline{\pmb{v}}^{\ast}\overline{m}$) and due to wave induced forces ($\pmb{F}$).} \begin{equation}
    \langle F_p\rangle_{\vartheta}=g^{-1}\int^{\frac{\pi}{2}}_{-\frac{\pi}{2}}\overline{\omega}^{\ast}\overline{m}\,a^{2}\cos\vartheta\,\text{d}\vartheta, 
\end{equation}
and the total meridional momentum flux associated with waves \begin{equation}
    \langle F^{\prime}_\vartheta\rangle_{p} = 2\pi a\cos\vartheta\int^{0}_{p_\text{s}} F^{\prime}_{\vartheta}\,\text{d}p/g, 
\end{equation}
respectively. $\omega^{\ast}$ is the residual pressure vertical velocity defined by (\ref{eq:res_circ}), and $F^{\prime}_{\vartheta}$ is the horizontal component of the EP flux defined by (\ref{eq:ep_flux}). $\langle F_p\rangle_{\vartheta}$ and $\langle F^{\prime}_\vartheta\rangle_{p}$ are shown averaged over various periods of the spin-up phase of the $\varOmega_{\text{E}}/16$ experiment. Solid lines present data from the 3D experiment, and dashed lines present data from the axisymmetric experiment. Blue lines show $\text{d}\overline{m}/\text{d}t$. 

During the axisymmetric phase, horizontal eddy angular momentum fluxes are small in the 3D experiment (\autoref{fig:v_mom_flux}, days 0-180), and the vertical flux of $m$ due to the overturning circulation in the 3D and axisymmetric experiments is virtually identical (\autoref{fig:w_mom_flux}, days 0-180); consequently, $\text{d}\overline{m}/\text{d}t$ in the free atmospheres of the two experiments is very similar during this period. At the end of the axisymmetric spin-up phase, \textsf{E}\,=\,\textsf{P}, and there is no net vertical transport of $m$ by the Hadley circulation. This is evident in \autoref{fig:w_mom_flux}, which shows $\langle F_p\rangle_{\vartheta}$ (the dashed orange curve) vanishing in the free atmosphere of the axisymmetric experiment after day 180.

When non-axisymmetric disturbances flux $m$ towards the equator, the downward transport of $m$ by the Hadley circulation, \textsf{P} in \autoref{fig:gierasch} will be reduced, but \textsf{E} will remain at its original value (the rising air is still equilibrated with the underlying equatorial surface). For as long as \textsf{P}\,<\,\textsf{E}, the net vertical transport of $m$ by the Hadley circulation will be upward, thus allowing $S$ in the 3D experiment to increase beyond that achieved in the axisymmetric experiment. Equatorward fluxes of $m$ due to wave induced forces in the 3D experiment become large during the period spanning days 180-360 (\autoref{fig:v_mom_flux}), and consequently the vertical flux of $m$ due to the residual mean circulation remains substantial in the 3D experiment during this period, contrasting the axisymmetric experiment where it has vanished (\autoref{fig:w_mom_flux}). At some point, air leaving the equator in the poleward branch will have sufficiently large $m$ that, after some angular momentum is returned to the equator via horizontal eddy transport, \textsf{P} will balance \textsf{E} once more. The system may also equilibriate if a downward vertical eddy angular momentum transport is established which balances upward transport by the mean flow. The equilibrated system has $S_{\text{3D}}>S_{\text{ax}}$. 

This mechanism for extracting angular momentum from the planetary surface and depositing it in the atmosphere was first proposed and demonstrated by \citet{1975JAtS...32.1038G}, and is often referred to as the Gierasch--Rossow--Williams mechanism. \citeauthor{1975JAtS...32.1038G} did not speculate on the nature of the non-axisymmetric disturbances which might induce substantial up-gradient eddy angular momentum transport, and instead parametrised eddy angular momentum transport with strong horizontal diffusion of angular velocity. \citet{1979JAtS...36..377R} later proposed that equatorward momentum transport induced by equatorial waves, generated by barotropic instability, could provide a realistic source of up-gradient transport. Recently it has been demonstrated that at low rotation rate, a planetary-scale barotropic, ageostrophic instability generates equatorial waves which can induce up-gradient angular momentum transport \citep{2014GeoRL..41.4118W,2018JAtS...75.2299Z}. 

\subsection{Reverse argument for rapidly rotating experiments}

The previous subsection has described how $S_{\text{3D}}$ may become larger than $S_{\text{ax}}$ for experiments run at low rotation rate. We have suggested that the establishment of up-gradient eddy angular momentum transport after an axisymmetric spin-up phase allows the Hadley circulation to achieve additional net upward vertical transport of angular momentum before the flow equilibritates. 

In principle, the same argument may be applied in \emph{reverse} for the rapidly rotating experiments to explain why $S_{\text{ax}}>S_{\text{3D}}$ (\autoref{fig:the_graph}). In the rapidly rotating three-dimensional experiments, the zonal flow in the sub-tropical (Hadley cell) region is decelerated by dissipating Rossby waves at the expense of acceleration in the mid-laitudes. This constitutes a down-gradient wave induced flux of angular momentum within the Hadley cell region (angular momentum is fluxed poleward). 

Returning to \autoref{fig:gierasch}, this will make the downward flux \textsf{P} greater than the upward flux \textsf{E}, and so the net effect of the overturning circulation during this phase will be to remove zonal angular momentum from the free atmosphere. As a result, $S_{\text{3D}}$ will be less than $S_{\text{ax}}$. In this scenario, the flow equilibrates when a net upward vertical eddy momentum transport balances the net downward transport by the mean flow within the overturning region.

\section{Scaling for global super-rotation in axisymmetric atmospheres}\label{sec:axissym}

In this section, we will develop a scaling theory for $S$, for the case of an axisymmetric, inviscid atmosphere. Then, in \autoref{sec:planets}, we will use the theory to interpret the high and low rotation rate behaviour of $S$ exhibited in our numerical experiments.

\subsection{The Held--Hou Model}

\citet[][hereafter \citetalias{1980JAtS...37..515H}]{1980JAtS...37..515H} present an analytic model for the circulation of an axisymmetric, inviscid atmosphere, derived from the Boussinesq hydrostatic primitive equations on a sphere. We will use the \citetalias{1980JAtS...37..515H} model to derive a scaling for $S$ in terms of the external thermal Rossby number $\mathcal{R}$. 

The \citetalias{1980JAtS...37..515H} model consists of essentially two layers, a lower layer with $u\approx0$, and an upper layer (the free atmosphere) where $u\left(\vartheta,z\right)\neq0$ and is in gradient wind balance with the model temperature field \begin{equation}
    \frac{\partial}{\partial z}\left(fu+\frac{u^{2}\tan\vartheta}{a}\right)=-\frac{g}{a\theta_{0}}\frac{\partial\theta}{\partial\vartheta},\label{eq:full_twb}
\end{equation}
where $g$ is the acceleration due to gravity, $\theta_0$ is the mean surface temperature, and $\theta$ is the potential temperature. The advective term which would appear on the left-hand-side of (\ref{eq:full_twb}) is assumed to be small and has been omitted. The atmospheric circulation is forced by a linear relaxation to a radiative-convective equilibrium potential temperature field $\theta_{\text{eq}}$ \begin{equation}
    \frac{\theta_{\text{eq}}}{\theta_0}=1-\frac{2}{3}\Delta_{h}P_{2}\left(\sin\vartheta\right)+\Delta_{v}\left(\frac{z}{H}-\frac{1}{2}\right)\label{eq:full_eq_T}
\end{equation}
where $\theta_0\Delta_h$ is the equator-to-pole surface temperature difference, and $P_2(x)=(3x^{2}-1)/2$ is the second Legendre polynomial.

(\ref{eq:full_twb}) and (\ref{eq:full_eq_T}) may be vertically integrated to yield \begin{equation}
    fu+\frac{u^{2}\tan\vartheta}{a}=-\frac{gH}{a\theta_0}\frac{\partial\overline{\theta}}{\partial\vartheta}\label{eq:twb}
\end{equation}
at $z=H$, and \begin{equation}
    \overline{\theta}_{\text{eq}}=\theta_{\text{0}}\left[1-\frac{2}{3}\Delta_h P_2(\sin\vartheta)\right], \label{eq:eq_T}
\end{equation}
where an overline denotes a vertical average $H^{-1}\int^{H}_{0}\text{d}z$, and $H$ is the domain height. 

In the free atmosphere, the circulation is divided into two regions: a tropical (or Hadley cell) region, and an extra-tropical region (as in \autoref{fig:had_schem}). In the extra-tropical region, $v=0$, so $u$ is determined through gradient wind balance by the radiative-convective equilibrium temperature field [i.e. $\theta=\theta_{\text{eq}}$ in (\ref{eq:full_twb})], and takes the form \begin{equation}
     u_\text{ET} = \varOmega a\cos\vartheta\left(\sqrt{2\mathcal{R}\frac{z}{H}+1} -1\right), \label{eq:u_e}
\end{equation}
where $\mathcal{R}\equiv(\Delta_h gH)/(\varOmega a)^{2}$ is the analogue of (\ref{eq:RoT}) for the Boussinesq equations [see Appendix A of \citet{2019JAtS...76.1397C} for a discussion regarding the correspondence between $\mathcal{R}$ in the Boussinesq and non-Boussinesq primitive equations]. 

In the Hadley cell region, $u$ is determined by conservation of specific axial angular momentum $m$. Air is assumed to rise at the equator with $u=0$, and $u$ depends on $\vartheta$ as \begin{equation}
    u_\text{HC}=\frac{\varOmega a\sin^{2}\vartheta}{\cos\vartheta}. \label{eq:ang_mom_u}
\end{equation}
The vertically averaged potential temperature field in the Hadley cell region is then determined to be in gradient wind balance with $u_{\text{HC}}$ [i.e. $u=u_{\text{HC}}$ in (\ref{eq:twb})]: \begin{equation}
    \frac{\overline{\theta}(0)-\overline{\theta}}{\theta_0} = \frac{u^{2}_{\text{HC}}}{2gH}.\label{eq:HC_T}
\end{equation}

\citetalias{1980JAtS...37..515H} then solve for the Hadley cell boundary latitude $\vartheta_{\text{H}}$ by enforcing two matching conditions: \begin{equation}
    \overline{\theta}\left(\vartheta_{\text{H}}^{+}\right) = \overline{\theta}\left(\vartheta_{\text{H}}^{-}\right), \label{eq:cont_t}
\end{equation}
and \begin{equation} 
    \int^{\vartheta_{\text{H}}}_{0}\overline{\theta}\cos\vartheta\,\text{d}\vartheta = \int^{\vartheta_{\text{H}}}_{0}\overline{\theta}_{\text{eq}}\cos\vartheta\,\text{d}\vartheta. \label{eq:consv_t}
\end{equation}
(\ref{eq:cont_t}) requires that potential temperature be continuous across the Hadley cell boundary, and (\ref{eq:consv_t}) requires that the Hadley cell is energetically closed. 

If (\ref{eq:eq_T}) and (\ref{eq:HC_T}) are substituted into (\ref{eq:consv_t}), then (\ref{eq:cont_t}) and (\ref{eq:consv_t}) may be used to derive the following expression \begin{equation}
    \mathcal{R} = \frac{3}{4}\left[\frac{1}{3}+\frac{1}{x_{\text{H}}^{2}}+\frac{x_{\text{H}}^{2}}{1-x^{2}_{\text{H}}}-\frac{1}{2x^{3}_{\text{H}}}\ln\left(\frac{1+x_{\text{H}}}{1-x_{\text{H}}}\right)\right] \label{eq:HHR}
\end{equation}
where $x_{\text{H}} = \sin\vartheta_{\text{H}}$, which may be solved numerically for $\vartheta_{\text{H}}$ given $\mathcal{R}$ (defined in terms of external parameters). 

\subsection{Expression for $S$} 

In order to relate $S$ to $\mathcal{R}$, we will make use of the zonal velocity profiles found for the tropical and extra-tropical regions of the \citetalias{1980JAtS...37..515H} model. We will assume \begin{equation}
    u=\begin{cases}u_{\text{HC}}, \ &\vartheta\in[0,\vartheta_{\text{H}}], \\ u_{\text{E}}, \ &\vartheta\in(\vartheta_{\text{H}},\pi/2]. \end{cases} \label{eq:u_scaling}
\end{equation}
$u_{\text{HC}}$ is given by (\ref{eq:ang_mom_u}) for all $p/p_s<0.8$ (i.e., in the free atmosphere), and $u_{\text{HC}}=0$ in the boundary layer. By assuming that $u_{\text{HC}}$ is independent of $z$ in the free atmosphere, we are assuming that $\partial\theta/\partial\vartheta\rightarrow 0$ within the Hadley cell. $u_{\text{ET}}$ is given by (\ref{eq:u_e}) through the depth of the atmosphere. 

The definition of $S$ may be re-written \begin{equation}
    S = \frac{\int \rho a\cos\vartheta\,u\,\text{d}V}{\int \rho \varOmega a^{2}\cos^{2}\vartheta\,\text{d}V}, \label{eq:glob_s2}
\end{equation}
as the $-1$ in (\ref{eq:glob_s}) cancels with the planetary ($\varOmega a^{2}\cos^{2}\vartheta$) contribution to $m$ in the numerator. $S$ may then be split into contributions from the Hadley cell and extra-tropical regions \begin{equation}
    S = S_{\text{HC}} + S_{\text{ET}} \label{eq:S2parts}
\end{equation}
where \begin{equation}
    S_{\text{HC}} = \frac{\int^{0}_{\frac{4}{5}p_s}\int_{0}^{\vartheta_{\text{HC}}} u_{\text{HC}}\,\cos^{2}\vartheta\,\text{d}\vartheta\,\text{d}p/g}{\int^{0}_{p_{s}}\int_{0}^{\frac{\pi}{2}}  \varOmega a\cos^{3}\vartheta\,\text{d}\vartheta\,\text{d}p/g},\label{eq:Shc}
\end{equation}    
and, \begin{equation}
    S_{\text{ET}} = \frac{\int^{H}_{0}\int_{\vartheta_{\text{HC}}}^{\frac{\pi}{2}}  u_{\text{E}}\,\cos^{2}\vartheta\,\text{d}\vartheta\,\rho_0\text{d}z}{\int^{H}_{0}\int_{0}^{\frac{\pi}{2}}   \varOmega a\cos^{3}\vartheta\,\text{d}\vartheta\,\rho_0\text{d}z}.\ \ \ \label{eq:Se}
\end{equation}
Note that in the Boussinesq framework, $\rho=\rho_0$ is a constant reference density, and we have used hydrostatic equilibrium $\text{d}p=-g\rho_0\text{d}z$ to write (\ref{eq:Shc}) as an integration over pressure. The integration $\int\text{d}\lambda$ in each expression cancels between the numerator and denominator due to axisymmetry, and is omitted. 

(\ref{eq:Shc}) and (\ref{eq:Se}) may then be evaluated analytically using (\ref{eq:ang_mom_u}) and (\ref{eq:u_e}) to obtain \begin{align}
    S &= S_{\text{HC}} + S_{\text{ET}} \nonumber \\ 
     &= \frac{2}{5}\sin^{3}\vartheta_{\text{H}}\ + \label{eq:S_scaling} \\ 
      & \quad\frac{1}{12}\left[\frac{\left(2\mathcal{R}+1\right)^{\frac{3}{2}}-1}{2\mathcal{R}}-\frac{3}{2}\right]\left(8-9\sin\vartheta_{\text{H}}-\sin3\vartheta_{\text{H}}\right). \nonumber  
\end{align}
As $\vartheta_{\text{H}}$ depends only on $\mathcal{R}$ [obtained by solving (\ref{eq:HHR})], $S$ is now determined solely by $\mathcal{R}$.

\begin{figure*}
    \centering\includegraphics[width=.90\linewidth]{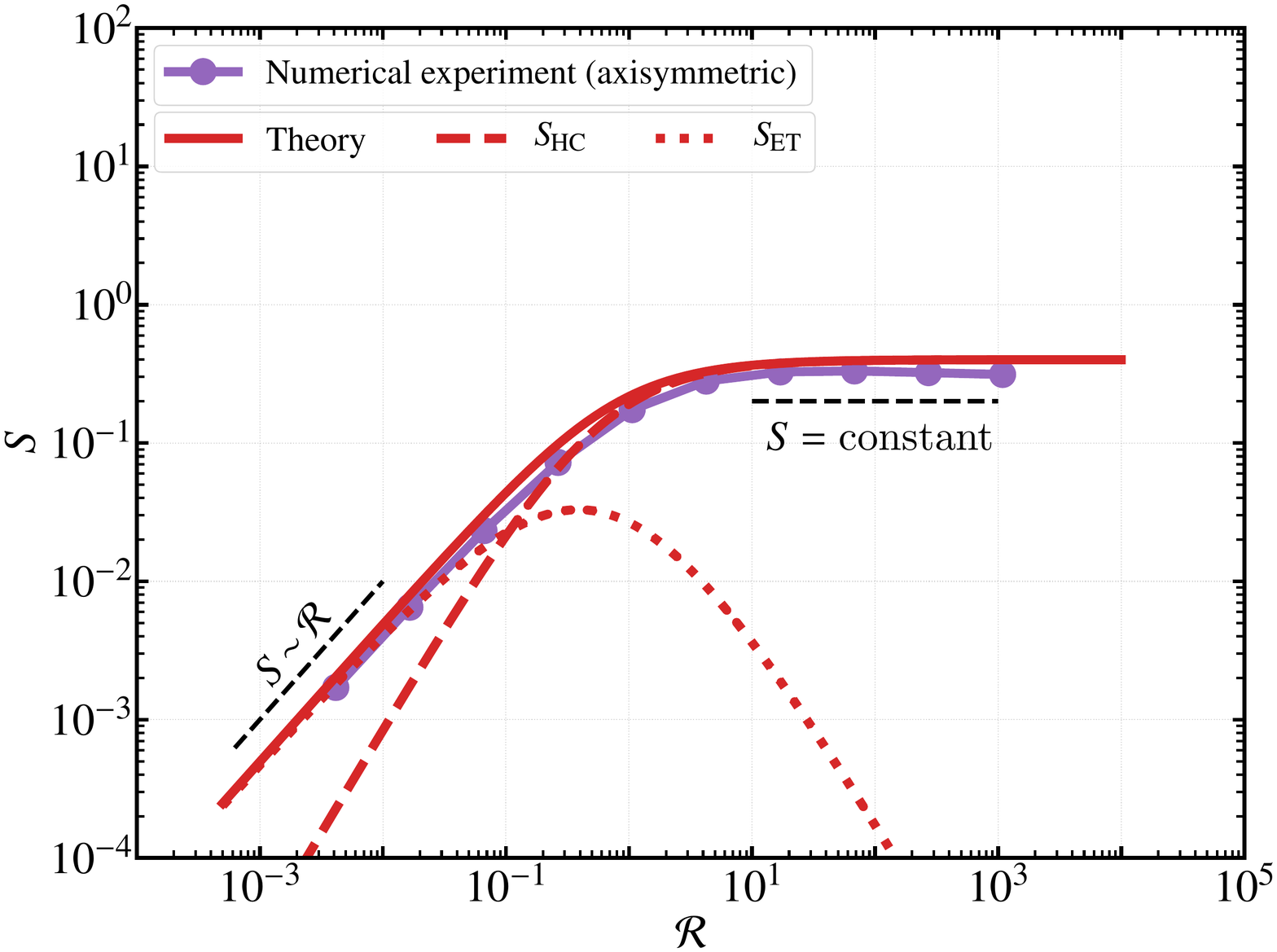}
    \caption{Theoretical prediction for $S$ from (\ref{eq:S_scaling}) vs. $\mathcal{R}$ for the case of an axisymmetric atmosphere (solid red curve). Also shown are the contributions to the theoretical $S$ from the Hadley cell and extra-tropical regions, $S_{\text{HC}}$ and $S_{\text{ET}}$. The purple line shows $S$ for the axisymmetric numerical experiments.}\label{fig:theory_graph}
\end{figure*}

\subsection{Comparison between theoretical $S$ and simulation results}

$S$ predicted by (\ref{eq:S_scaling}) is plotted against $\mathcal{R}$ in \autoref{fig:theory_graph}. The values for $S$ calculated for the axisymmetric numerical experiments are also shown, and there is good agreement between the theoretical prediction for $S$ and the numerical experiments in both the high and low rotation rate limits. 

(\ref{eq:S_scaling}) is comprised of two terms, one that measures the contribution to $S$ from the Hadley cell region ($S_{\text{HC}}$), and one that measures the contribution from the extra-tropical region ($S_{\text{ET}}$). These are shown in \autoref{fig:theory_graph} as a dashed and dotted curve, respectively. At high rotation rates, the extra-tropical term dominates the total value of $S$, as in this regime the Hadley Cell is small and the `extra-tropics' essentially comprise the whole atmosphere. At low rotation rates, the contribution to $S$ from the Hadley cell dominates, as the Hadley cell expands to fill the domain. 

The transition between the extra-tropical and Hadley-cell dominated regimes occurs when the first and second terms in (\ref{eq:S_scaling}) are equal. This occurs when $\mathcal{R}=0.11$, corresponding to a Hadley cell latitude $\vartheta_{\text{H}}=22.8^{\circ}$.

\section{Regimes of global super-rotation}\label{sec:planets}

\begin{figure*} 
    \centering\includegraphics[width=.90\linewidth]{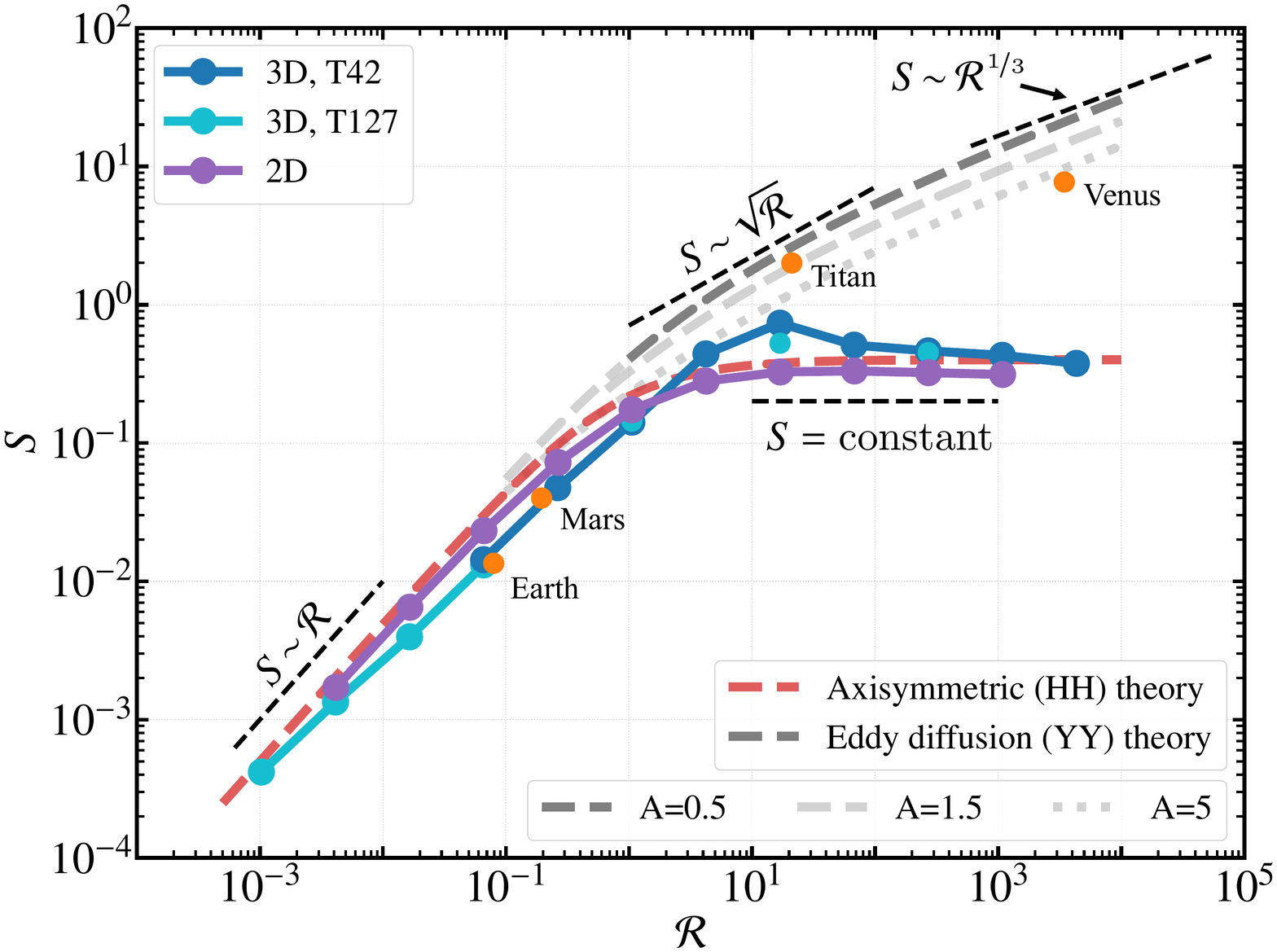}
    \caption{$S$ vs. $\mathcal{R}$ calculated from numerical experiments (solid curves) and theoretical predictions (dashed curves). 3D experiments are shown in blue, and axisymmetric experiments in purple (see \autoref{fig:the_graph} caption for details). The orange dots show estimates for $S$ for the Earth, Mars, Titan and Venus taken from \citet{2018AREPS..46..175R}. The red dashed curve shows $S$ predicted by the axisymmetric theory (\ref{eq:S_scaling}). The grey dashed curves show $S$ predicted by the quasi-axisymmetric theory (\ref{eq:SYY}) of \citet{YAMAMOTOYODEN2013}, where the parameter values for $A$ are shown in the legend (see text for definition).}\label{fig:theory_graph2}
\end{figure*}

In this section we will discuss different scaling regimes for global super-rotation. \autoref{fig:theory_graph2} compiles the information shown in \autoref{fig:the_graph} and \autoref{fig:theory_graph}. Values for $S$ estimated for the real planets, calculated from our axisymmetric and three-dimensional experiments, and predicted by the axisymmetric theory are shown. Theoretical predictions for $S$ in the $\mathcal{R}\gg1$ regime proposed by \citet[][discussed later in this section]{YAMAMOTOYODEN2013} are also presented.

\subsection{Geostrophic regime\ \ ($\mathcal{R}\ll1$)}

It is instructive to consider the behaviour of the axisymmetric theory in the limits $\mathcal{R}\ll1$ and $\mathcal{R}\gg1$. When $\mathcal{R}\ll1$ (rapid rotation) \begin{equation}
    \frac{\left(2\mathcal{R}+1\right)^{\frac{3}{2}}-1}{2\mathcal{R}}=\frac{3}{2}+\frac{3\mathcal{R}}{4}+\dots, 
\end{equation}
(via Taylor expansion) so that \begin{equation}
    \frac{\left(2\mathcal{R}+1\right)^{\frac{3}{2}}-1}{2\mathcal{R}} - \frac{3}{2} \approx \frac{3\mathcal{R}}{4}. 
\end{equation}
Additionally $\vartheta_{\text{H}}\rightarrow0$, meaning that $\sin\vartheta_{\text{H}}$, $\sin3\vartheta_{\text{H}}$, and $\sin^{3}\vartheta_{\text{H}}\ \rightarrow0$. We then have \begin{equation}
    S\approx \frac{1}{2}\mathcal{R}. \label{eq:smallR_scaling}
\end{equation}
In the rapidly rotating regime, $S$ is dominated by the extra-tropical region (as the Hadley Cell is small), and (\ref{eq:smallR_scaling}) predicts that $S$ scales linearly with $\mathcal{R}$, which corresponds to a $\varOmega^{-2}$ scaling with rotation rate. 

The $\mathcal{R}$ dependence in (\ref{eq:S_scaling}) comes from the expression for $u_{\text{ET}}$, (\ref{eq:u_e}), which originally derives from enforcing gradient wind balance in the extra-tropical region with the radiative-convective equilibrium forcing profile. This tells us that the $S\propto\varOmega^{-2}$ scaling in the rapidly rotating regime is simply that implied by geostropic thermal wind balance. If we write $u\,{\sim}\,U$, $f\,{\sim}\,\varOmega$, and $\partial\overline{\theta}/\partial\vartheta\,{\sim}\,\Delta\theta$, then the gradient wind equation (\ref{eq:twb}) is approximately \begin{equation}
    \varOmega U\ \pm\ \frac{U^{2}}{a}\ \sim\ \frac{gH\Delta\theta}{a\theta_0}. \label{eq:scale_analysis}
\end{equation}
If we assume $S\,{\sim}\,U/(\varOmega a)$, then in the rapidly rotating regime where $\varOmega U\gg U^{2}/a$ (i.e., geostrophic balance holds), and $\Delta\theta/\theta_0\rightarrow\Delta_h$,  \begin{equation}
    U \sim \frac{gH\Delta_h}{\varOmega a};\qquad S\sim \mathcal{R}. 
\end{equation}
Thus, we refer to the rapidly rotating regime as the \emph{geostrophic regime}. 

In the rapidly rotating (small $\mathcal{R}$) regime, non-axisymmetric disturbances have a limited effect on the zonal angular momentum budget in the 3D experiments, and the difference in $S$ between the 3D and axisymmetric experiments is small. This means that the axisymmetric theory (\ref{eq:S_scaling}) closely estimates $S$ not only for the axisymmetric experiments, but also for the 3D experiments. 

The values for $S$ estimated for the Earth and Mars by \citet{2018AREPS..46..175R}, $S_{\text{E}}=0.014$ and $S_{\text{M}}=0.04$, are close to those obtained in the 3D numerical experiments, and both planets appear to lie in the geostrophic regime identified by the axisymmetric theory. This is certainly the case for the Earth, but for Mars we note that its external Rossby number $\mathcal{R}_{\text{M}}=0.19$ is slightly larger than the critical $\mathcal{R}=0.11$ where the Hadley cell and extra-tropical terms contributing to $S$ in (\ref{eq:S_scaling}) become of comparable size.

\subsection{Angular momentum conserving regime ($\mathcal{R}\gg1$, axisymmetric or nearly axisymmetric inviscid flow)}

When $\mathcal{R}\gg1$ (slow rotation) then $\vartheta_{\text{H}}\rightarrow\pi/2$, so that $\sin\vartheta_{\text{H}}$ and $\sin^{3}\vartheta_{\text{H}} \rightarrow 1$, while $\sin3\vartheta_{\text{H}}\rightarrow-1$. In this scenario, the axisymmetric theory approaches \begin{equation}
    S\approx\frac{2}{5}=\text{const}. 
\end{equation}
The $\mathcal{R}\gg1$ limit is the scenario where the Hadley cell has filled the entire domain. This is qualitatively the same as the $S=1/2$ limit suggested in \autoref{sec:hide_S}, where the entire atmosphere has specific zonal angular momentum $m=\varOmega a^{2}$ due to the equilibration of air with the equatorial surface, and conservation of $m$ in the free atmosphere. The additional factor $4/5$ comes from assuming $u=0$ in the boundary layer ($p/p_{\text{s}}>0.8$). If we write $u\,{\sim}\,U$ once more and apply this scaling to the definition of $u_{\text{HC}}$ we obtain \begin{equation}
    U\sim\varOmega a;\qquad S\sim1=\text{const}.  
\end{equation}
$S$ measures $m$ relative to the planetary angular momentum, and thus is constant in the slowly rotating $\mathcal{R}\gg1$ regime. We will refer to the $S=\text{const.}$ regime as the \emph{angular momentum conserving regime}.

The $S=\text{const.}$ scaling may also be inferred from the gradient wind equation. In our slowly rotating numerical experiments, meridional heat transport by the (global) Hadley circulation effectively eliminates meridional temperature contrasts so that $\Delta\theta\rightarrow0$. Then (\ref{eq:scale_analysis}) implies \begin{equation}
    \varOmega U \sim \frac{U^{2}}{a};\qquad S \sim1.
\end{equation}

In the 3D experiments, the effect of non-axisymmetric disturbances on the zonal angular momentum budget becomes more pronounced when $\mathcal{R}$ is made large. Wave induced up-gradient transport of $m$ allows the mean meridional overturning to extract more angular momentum from the planetary surface than is achieved in the axisymmetric experiments, and $S_{3\text{D}}>S_{\text{ax}}$. In spite of this, the 3D experiments remain relatively close to the axisymmetric experiments, and when $\mathcal{R}$ is made  larger still (e.g. $\mathcal{R}>100$) the difference $S_{\text{3D}}-S_{\text{ax}}$ actually appears to reduce as $\mathcal{R}$ is increased. In this limit the 3D experiments approach the angular momentum conserving regime occupied by $S_{\text{ax}}$. 

It is not obvious that $S$ should stop increasing when $\mathcal{R}$ is made very large; global super-rotation much stronger than that achieved in our 3D experiments \emph{is possible} in an atmosphere with non-axisymmetric disturbances, as evidenced by the atmospheres of Venus and Titan. Analysis of the barotropic eddy kintetic energy power spectra for the 3D experiments (not shown) reveals a spectrum peaked at global wavenumber one for the $\varOmega_{\text{E}}/8$ and $\varOmega_{\text{E}}/16$ experiments. At even lower rotation rates (larger $\mathcal{R}$), the power spectra become flatter, eddy kinetic energy is more evenly distributed across scales, and local super-rotation is accelerated by instabilities with a zonal number greater than $1$ (i.e., shorter wavelength disturbances). It may be the case that the primary instability responsible for accelerating super-rotation at the equator (in the $\varOmega_{\text{E}}/8$ and $\varOmega_{\text{E}}/16$ experiments) has a short-wavelength cut-off that becomes larger than the domain size when $\mathcal{R}$ is increased appreciably, after which local super-rotation is accelerated by secondary instabilities which are less effective at inducing up-gradient transport of $m$. This argument is similar to that made to explain transition into an axisymmetric regime in the rotating cylindrical annulus, when the ``Eady short-wavelength cut-off'' for baroclinic instability is made larger some critical size \citep{1975AdPhy..24...47H}.

\subsection{Cyclostrophic regime ($\mathcal{R}\gg1$, non-axisymmetric or viscous flow)}

The degree of global super-rotation obtained in the three-dimensional numerical experiments when $\mathcal{R}\gg1$ is much lower than the estimates of $S$ for Venus and Titan calculated by \citet{2018AREPS..46..175R}. 

At low rotation rate, the three-dimensional numerical experiments reside close to the angular momentum conserving regime predicted by our axisymmetric theory. This regime is consistent with a scale analysis of the gradient wind balance equation (\ref{eq:twb}) when \begin{equation}
    U\gg\frac{gH\Delta\theta}{\varOmega a\theta_0}. \label{eq:inequality}
\end{equation}
This inequality is satisfied in our numerical experiments as $\Delta\theta\,{\rightarrow}\,0$ within the Hadley cell, which fills the domain at low rotation rate. If we let $gH\Delta\theta/\theta_0\approx R_{\text{d}}\Delta T$, then we may estimate the inequality (\ref{eq:inequality}) for Venus and Titan. For Venus, we assume $U\,{\sim}\,100\,\text{m\,s}^{-1}$. $\varOmega$ and $a$ are set to the values given in \autoref{tab:RoT}, and $\Delta T=10\,\text{K}$ is used as the actual (i.e., not the radiative-convective equilibrium) equator-to-pole temperature contrast. Then $R_{\text{d}}\Delta T/(\varOmega a)\,{\sim}\,1000\,\text{m\,s}^{-1}$ for Venus. Similarly for Titan, $U\,{\sim}\,100\,\text{m\,s}^{-1}$ and $R_{\text{d}}\Delta T/(\varOmega a)\,{\sim}\,200\,\text{m\,s}^{-1}$. Therefore the inequality (\ref{eq:inequality}) is not satisfied for either planet.

Venus and Titan may instead occupy a regime where global super-rotation scales cyclostrophically. If we assume $\varOmega U \ll U^{2}/a$, then a scale analysis of the gradient wind equation suggets that the local centrifugal term balances the pressure gradient term, and (\ref{eq:scale_analysis}) becomes 
\begin{equation}
    U^{2} \sim \frac{gH\Delta\theta}{\theta_0};\qquad S \sim \sqrt{\frac{gH\Delta\theta/\theta_0}{(\varOmega a)^{2}}}\approx\sqrt{\mathcal{R}}. \label{eq:cyclo} 
\end{equation}
The final `$\approx$' appears by assuming $\Delta\theta/\theta_0\approx\Delta_h$. $S\,{\sim}\,\sqrt{\mathcal{\mathcal{R}}}$ is shown as a black dashed line in \autoref{fig:theory_graph2}. The cyclostrophic scaling appears to loosely approximate the slope $\Delta S_{\text{TV}}=(S_{\text{V}}-S_{\text{T}})/(\mathcal{R}_{\text{V}}-\mathcal{R}_{\text{T}})$, although it is clearly steeper than $\Delta S_{\text{TV}}$. 

As $S$ in Venus' and Titan's atmospheres greatly exceeds the upper bound $S=1/2$ achievable in an axisymmetric atmosphere (see \autoref{sec:hide_S}), substantial up-gradient eddy angular momentum transports are clearly important in generating Venus' and Titan's strong global super-rotation. Our axisymmetric theory is therefore unable to predict the correct magnitude for $S$ estimated for their atmospheres; in order to predict $S$ correctly for Venus and Titan, some parametrisation of up-gradient eddy angular momentum transport would need to be included in our model. 

\citet[][hereafter \citetalias{YAMAMOTOYODEN2013}]{YAMAMOTOYODEN2013} present a theoretical model for a non-dimensional measure of super-rotation strength, $S^{\prime}$ defined as \begin{equation}
    S^{\prime}\equiv\frac{U}{\varOmega a}, 
\end{equation}
where $U$ is a scale for the zonal velocity at the model top. In \autoref{ap:SYY} we show that $S^{\prime}$ is related to $S$ by $S^{\prime}=4S/3$ for the theoretical model presented in \citetalias{YAMAMOTOYODEN2013}.

The model is derived from the Boussinesq primitive equations on a sphere forced by Newtonian relaxation (e.g., as in \citetalias{1980JAtS...37..515H} and in this work), in the presence of diffusion of angular velocity. The diffusion terms are formulated in analogy with molecular viscosity, and can transport angular momentum up-gradient \citep{1986QJRMS.112..253R}. There is no heat transport associated with the diffusion parametrisation. In order to obtain an expression for $S^{\prime}$ in terms of external parameters, \citetalias{YAMAMOTOYODEN2013} assume that the relaxation time for horizontal diffusion is much shorter than (i) the turnover time of the meridional circulation, and (ii) the relaxation time for vertical diffusion. As horizontal diffusion is assumed to be large, \citetalias{YAMAMOTOYODEN2013} set the zonal velocity at the model top to take the form of solid-body rotation, $u\propto\cos\vartheta$. It has been shown that \citep{YAMAMOTOYODEN2009} meridional overturning takes the form of a single equator-to-pole Hadley cell in the presence of large horizontal diffusion, which motivates \citetalias{YAMAMOTOYODEN2013} to set $v\propto\pm\sin2\vartheta$ at the top $(+)$ and bottom $(-)$ of the domain. The assumptions (i) and (ii) mean that the \citetalias{YAMAMOTOYODEN2013} theory does not have an inviscid limit.

Under the assumptions described in the previous paragraph, \citetalias{YAMAMOTOYODEN2013} derive a theoretical relationship between $S^{\prime} = 4S/3$, and external parameters $A$, $B$ and $\mathcal{R}$, \begin{equation}
    \left[S^{\prime2}+2S^{\prime}+BS^{\prime}\left(\frac{2+S^{\prime}}{1+S^{\prime}}\right)\right]\left[\frac{AS^{\prime}}{2}\left(\frac{2+S^{\prime}}{1+S^{\prime}}\right)+1\right]=2\mathcal{R}, \label{eq:SYY}
\end{equation}
where $\mathcal{R}$ is the same as in our axisymmetric theory, $A=\pi^{2}\tau/\tau_{\text{V}}$ is proportional to the ratio of the radiative relaxation time $\tau$ to the timescale for vertical eddy momentum diffusion $\tau_{\text{V}}\equiv H^{2}/\nu_{\text{V}}$, and $B=20\pi^{2}\left(\varOmega^{-1}/\sqrt{\tau_{\text{H}}\tau_{\text{V}}}\right)^{2}$ is proportional to the square of the ratio between the planetary rotation period and the geometric mean of the timescales for horizontal and vertical eddy momentum diffusion, $\tau_{\text{H}}\equiv a^{2}/\nu_{\text{H}}$ and $\tau_{\text{V}}$.  

$S=3S^{\prime}/4$ predicted by (\ref{eq:SYY}) is shown in \autoref{fig:theory_graph2} as a series of grey dashed curves (different curves correspond to different values for $A$) plotted against $\mathcal{R}$. We only show predictions for $S$ from (\ref{eq:SYY}) for $\mathcal{R}>0.1$, as when $\mathcal{R}\ll1$ it is generally no longer appropriate to assume up-gradient eddy angular momentum transport (see discussion in Section 5.5 of \citetalias{YAMAMOTOYODEN2013}). The parameter $B$ in (\ref{eq:SYY}) is only important in determining the behaviour of $S^{\prime}$ at low and intermediate $\mathcal{R}$ (high and intermediate rotation rate), and has no affect on the $\mathcal{R}\gg1$ asymptote. If $B\gg1$, then the geostrophic scaling for $S$ at low $\mathcal{R}$ is replaced by a scaling determined by `horizontal diffusion balance' between the horizontal diffusion and pressure gradient terms in the horizontal momentum equations. For all of the grey \citetalias{YAMAMOTOYODEN2013} curves shown in \autoref{fig:theory_graph2} we choose $B=0.3$. $B<1$ puts the curves in the geostrophic regime occupied by our numerical experiments, the Earth, and Mars, and the specific value $0.3$ is chosen so that the \citetalias{YAMAMOTOYODEN2013} theory curves join up with our axisymmetric theory at approximately $\mathcal{R}=0.1$. The values for $A$ shown in \autoref{fig:theory_graph2} are chosen to be in the `correct' ballpark for Venus and Titan (see Section 5.5 of \citetalias{YAMAMOTOYODEN2013}), and multiple values are only shown to demonstrate the dependence of $S^{\prime}$ on $A$. $A$ is made larger when the coefficient for vertical eddy momentum diffusion (which transports angular momentum down-gradient)  is made larger, and increasing $A$ weakens the super-rotation strength.

\citetalias{YAMAMOTOYODEN2013} show that for intermediate values for $\mathcal{R}$, such that $1<\mathcal{R}<10$, $S^{\prime}$ scales with $\sqrt{2\mathcal{R}}\propto\varOmega^{-1}$, and that this may be interpreted as either (i) a cyclostrophic scaling similar to (\ref{eq:cyclo}), where the equilibrium $\Delta T$ is close to the radiative-convective equilibirum $\Delta T_{\text{eq}}$, or (ii) a modified-geostrophic scaling where $\Delta T$ is significantly reduced with respect to $\Delta T_{\text{eq}}$, which causes the gesotrophic scaling to change from ${\sim}\,\mathcal{R}$ to ${\sim}\,\sqrt{\mathcal{R}}$. Regime (i) corresponds to $A\ll1$, while (ii) corresponds to $A\gg1$. For large $\mathcal{R}>10$, a third (iii) regime exists where $S^{\prime}$ scales with $\mathcal{R}^{\nicefrac{1}{3}}\propto\varOmega^{-\nicefrac{2}{3}}$. This is a modified-cyclostrophic regime. The $\mathcal{R}^{\nicefrac{1}{3}}$ scaling differs from our $\sqrt{\mathcal{R}}$ estimate (\ref{eq:cyclo}) as $\Delta T$ is significantly reduced with respect to $\Delta T_{\text{eq}}$, and so $\sqrt{gH\Delta\theta/(\theta_0^{\nicefrac{1}{2}}\varOmega a)^{2}}\,{\not\approx}\,\sqrt{\mathcal{R}}$.

Inspection of \autoref{fig:theory_graph2} hints at the possibility that the $\varOmega_{\text{E}}/8$ and $\varOmega_{\text{E}}/16$ 3D experiments (particularly at T42 resolution) enter either the modified-geostrophic or cyclostrophic regime, before the primary instability accelerating super-rotation collapses as the rotation rate is reduced further. Titan may occupy any of the three regimes; acknowledging uncertainty in our estimates for $S_{\text{T}}$ and $\mathcal{R}_{\text{T}}$, however, we cannot be specific about whether Titan occupies regime (i), (ii) or (iii). \citetalias{YAMAMOTOYODEN2013} estimate $A=10^{1}\,{-}\,10^{2}$ for Titan, which would place it in the geostrophic regime (ii) if $\mathcal{R}$ is $\mathcal{O}(10)$. To obtain this estimate, however, they `guess' $E_{\text{v}}=\nu_{\text{v}}/(\varOmega H^{2})=10^{-3}$, which is poorly constrained as $\nu_{\text{V}}$ parametrises the effect of non-axisymmetric disturbances for which observations are sparse. If our estimate for $\mathcal{R}_{\text{T}}$ is an underestimate, then Titan may actually occupy regime (iii). Venus' atmospheric circulation likely occupies regime (iii). 

As the region $1<\mathcal{R}<10$ is a small component of the parameter space we consider in this work, for the remainder of this manuscript we will refer to the set of regimes identified by \citetalias{YAMAMOTOYODEN2013} in the $\mathcal{R}\gg1$ limit as the \emph{cyclostrophic regime}.

\subsection{A `default' regime for slowly rotating planets?}

The emergence of strong super-rotation is often assocaited with low planetary rotation rate, or more generally, large $\mathcal{R}$. This is partly due to the existence of strong global super-rotation in the atmospheres of Venus and Titan, and additionally because local super-rotation has been shown to emerge in idealised `Earth-like' numerical models when $\mathcal{R}$ is made large [e.g. \citet{2010JGRE..11512008M}]. Our numerical experiments have shown, however, that the degree of global super-rotation in an Earth-like atmosphere may saturate before reaching a strength comparable to that in Venus' and Titan's atmospheres. In this scenario, the atmospheric circulation enters an angular momentum conserving regime, as opposed to the cyclostrophic regime occupied by Venus and Titan. This motivates the question: which regime is \emph{typical} of a slowly rotating planet? 

In our numerical experiments, the atmospheric circulation is forced by a linear relaxation to a \emph{time-independent, axisymmetric} radiative-convective equilibrium temperature profile. Our numerical model is not configured to realistically simulate the atmosphere of either Venus or Titan, and thus is likely missing features specific to their atmospheres that allow them to enter the cyclostrophic regime. For example, in the case of Venus, strong global super-rotation is likely maintained at least partially by the diurnal and semi-diurnal tides excited due to its long solar day, and the existence of substances in the atmosphere that absorb at UV wavelengths \citep{2017SSRv..212.1541S}. In the case of Titan, the source of strong super-rotation is less clear, although it may be related \citep{2006JAtS...63.1548W} to the fact that Titan's stratosphere, where super-rotation is strongest, is strongly statically stable \citep{2005Natur.438..785F,2009RSPTA.367..649F}. Neither of these properties are represented in our numerical experiments. 

One conclusion that may be drawn from our simulation results is that super-rotation of a strength comparable to that of Venus and Titan \emph{cannot} be induced solely by reducing the planetary rotation rate in an atmospheric model forced by diabatic heating with an Earth-like vertical structure and no diurnal cycle. We would therefore argue that the `default' regime for a slowly rotating planet to occupy is the angular momentum conserving regime, which exhibits mild global (and local) super-rotation $S\,{\sim}\,1$. 

\subsection{Tidally-locked planets}

In this work, we have not considered the atmospheres of tidally-locked planets, which permanently present the same fact to their host star. Theoretical work and numerical modelling has shown that the atmospheres of tidally-locked planets may often feature super-rotation in their atmospheres -- driven in direct response to the stationary heating associated with their orbital configuration [see \citet{2019AnRFM..51..275P} for a review]. It would be an interesting topic for future work to investigate how $S$ depends on $\varOmega$ in the atmospheres of tidally-locked terrestrial planets. The up-gradient angular momentum transport associated with the stationary heating may be sufficiently large to allow tidally-locked planets to access the cyclostrophic scaling for $S$ at low rotation rate. The relation between $S$ in atmospheres with axisymmetric forcing and tidally-locked atmospheres may be complicated, however, as the overturning circulation in tidally-locked atmospheres also takes a different form in response to the stationary heating (i.e., it can no longer be described as an axisymmetric meridional overturning cell).

\section{Summary}\label{sec:conclude}

We have explored the rotation rate senstivity of terrestrial atmospheric circulations with an idealised numerical model -- run in both an axisymmetric and a three-dimensional configuration. The aim of our investigation was to reveal how global super-rotation depends on planetary rotation rate. 

The atmospheric circulation in each of our experiments is forced by linear relaxation to a zonally symmetric radiative-convective equilibrium temperature profile, which establishes a meridional overturning circulation. In the axisymmetric experiments, conservation of zonal angular momentum within the overturning circulation leads to the generation of sub-tropical zonal jets. As there is no mechanism for up-gradient angular momentum transport in an axisymmetric, inviscid atmosphere \citep{1969JAtS...26..841H}, local super-rotation is not present in any of the axisymmetric experiments. In the three-dimensional experiments, the atmospheric circulation is modified by eddy induced fluxes of heat and angular momentum. At rapid rotation rates, this leads to the generation of mid-latitude eddy-driven jets poleward of the sub-tropical jets associated with angular momentum conservation in the Hadley circulation. At slow rotation rates, eddies induce up-gradient angular momentum fluxes within the sub-tropical region, leading to the generation of local super-rotation.  

For each of our experiments we have calculated the global super-rotation index $S$ which compares the total axial angular momentum of an atmosphere with that achieved by solid-body co-rotation with the underlying planet. At high rotation rate, our experiments occupy a \emph{geostrophic} regime, where $S\,{\sim}\,\mathcal{R}\propto\varOmega^{-2}$. At low rotation rate, our experiments enter an \emph{angular momentum conserving} regime, where $S=\text{const}$. Both of these regimes are captured by an axisymmetric theoretical model for $S$ derived from the \citetalias{1980JAtS...37..515H} model, and the scaling for $S$ in each regime can also be obtained by conducting a scale analysis on the thermal wind equation. 

We note that $S$ achieved by our three-dimensional experiments remains close to that in the axisymmetric experiments, with some differences in detail. At high rotation rate, $S_{\text{3D}}<S_{\text{ax}}$, and at low rotation rate $S_{\text{3D}}>S_{\text{ax}}$. We show that $S_{\text{3D}}$ is greater than $S_{\text{ax}}$ at low rotation rate because non-axisymmetric disturbances allow for enhanced upward transport of angular momentum by the mean overtuning circulation during the model spin-up phase. We suggest that the same argument may be applied in reverse to explain why $S_{\text{3D}}<S_{\text{ax}}$ at high rotation rate. 

Estimates of $S$ for the Earth and Mars agree closely with those calculated from our numerical experiments. We conclude that the Earth's atmosphere is in the geostrophic regime, and Mars' atmosphere sits at the transition between the geostrophic and angular momentum conserving regimes. The atmospheres of Venus and Titan exhibit global super-rotation far stronger than that achieved by our numerical experiments and predicted by our axisymmetric theory. We find that the quasi-axisymmetric theory for $S$ presented in \citet[][\citetalias{YAMAMOTOYODEN2013}]{YAMAMOTOYODEN2013} is able to better predict the value for $S$ estimated for their atmospheres. Venus' atmosphere resides within a modified-cyclostrophic regime where $S\,{\sim}\,\mathcal{R}^{\nicefrac{1}{3}}\propto\varOmega^{-\nicefrac{2}{3}}$, and Titan's atmosphere may either in this regime, or a regime where $S\,{\sim}\,\mathcal{R}^{\nicefrac{1}{2}}\propto\varOmega^{-1}$ (modified-geostrophic or cyclostrophic). 

In the \citetalias{YAMAMOTOYODEN2013} theory, the cyclostrophic regime is accessed by requiring that there is large horizonal eddy diffusion, which is associated with substantial up-gradient angular momentum transport. Simply reducing the rotation rate in our idealised three-dimensional numerical model is not sufficient to induce up-gradient eddy angular momentum fluxes of the magnitude required to enter the cyclostrophic regime. We therefore argue that the `default' regime for a slowly rotating planet is the angular momentum conserving regime, characterised by mild global super-rotation. An important topic for future research is to determine which planetary parameters and processes (e.g. static stability, solar day length) are most important in causing the atmospheric circulation to transition between the angular momentum conserving and cyclostrophic regimes at low rotation rate.

\section*{Acknowledgements}

We would like to thank Mark Hammond for engaging in discussions which benefited this work, and Man-Suen Chan for IT support. We are grateful to all those who have contributed to the development of the \texttt{Isca} modelling framework (see \url{https://ex.ac.uk/isca}). The authors were supported by STFC grants ST/S505638/1, ST/S000461/1 and ST/N00082X/1.

\appendix
\setcounter{equation}{0}
\renewcommand{\theequation}{\thesection\arabic{equation}}
\section{Relation between global $S$ and local $s$} \label{ap:gregs}

In the definition of $S$, \eqref{eq:glob_s}, the numerator and denominator of the quotient are respectively the total atmospheric zonal angular momentum $M$,
and the contribution due to the planetary rotation, $P$. The other contribution to $M=P+W$ is due to the zonal wind: $W\equiv\int\rho ua\cos\vartheta\,\text{d}V$. Thus the global super-rotation index
\begin{equation}
S = \frac{M}{P} - 1 = \frac{W}{P} = \frac{\int \omega\,\text{d}I}{\int \varOmega\,\text{d}I} = \frac{\int c\,\text{d}I}{\int \text{d}I}.
\end{equation}
Here $\text{d}I = \rho a^2\cos^2\vartheta\, \text{d}V$ is an element of axial moment of inertia. 
$S$ is therefore a moment-of-inertia-weighted average of $c\equiv\omega/\varOmega$, where $\omega=u/(a\cos\vartheta)$ is the angular velocity of the zonal wind around the planet's axis. $c$, or $c/2$, can be thought of as the local `cyclostrophicity' --- $c/2$ being the ratio of the cyclostrophic to the geostrophic term in the thermal wind equation (\ref{eq:twb}). $c$ and $S$ can also be thought of as internal Rossby numbers (whereas $\mathcal{R}$ is an external Rossby number).

$c$ is related to the local super-rotation index $s$ via
\begin{equation}
m = (s+1)\varOmega a^2 = (c+1)\varOmega a^2\cos^2\vartheta
\end{equation}
and an expression analogous to \eqref{eq:loc_s} may be written
\begin{equation}
c = \frac{m}{\varOmega a^2\cos^2\vartheta} - 1.
\end{equation}
$c$ is a measure of $m$ relative to the planetary specific zonal angular momentum (the denominator) locally, whereas $s$ is relative to the value at the equator $\vartheta=0$.

A global super-rotation measure defined naturally in terms of $s$ would be
\begin{equation}
{\mathcal{S}} \equiv \frac{\int \rho m\,\text{d}V}{\int\rho \varOmega a^2\,\text{d}V}-1=\frac{\int s\rho\,\text{d}V}{\int\rho\,\text{d}V} \label{eq:a4}
\end{equation}
$\mathcal{S}$ is the mass-weighted average of $s$. Let the first denominator here, replacing $P$ in  \eqref{eq:glob_s}, be denoted $\mathcal{P}=\mathcal{M}\varOmega a^2$ (where $\mathcal{M}$ is the total atmospheric mass). $\mathcal{S}$ is related to $S$ via
\begin{equation}
M = (\mathcal{S}+1)\mathcal{P} = (S+1)P. \label{eq:a5}
\end{equation}
Furthermore, in the case of constant surface pressure, $P=2\mathcal{P}/3$, and so \begin{equation}
    \mathcal{S}[s]=(2S-1)/3. \label{eq:a6}
\end{equation}%
This equation, or more generally (\ref{eq:a5}), relates $S$ to $s$ via (\ref{eq:a4}). 

If $s = 0$ everywhere, then $\mathcal{S} = 0$ and $S = 1/2$ (as described in \autoref{sec:hide_S}). $\mathcal{S}>0$ requires the existence of processes that can generate up-gradient transfer of $m$.

\setcounter{equation}{0}
\renewcommand{\theequation}{\thesection\arabic{equation}}
\section{Relation between $S$ and $S^{\prime}$ for \citetalias{YAMAMOTOYODEN2013} model}\label{ap:SYY}

\citet[][\citetalias{YAMAMOTOYODEN2013}]{YAMAMOTOYODEN2013} define their non-dimensional measure of super-rotation strength $S^{\prime}$ to be \begin{equation}
    S^{\prime}\equiv\frac{U}{\varOmega a}, 
\end{equation}
where $U$ is a scale for the zonal velocity at the model top. $S^{\prime}$, like $S$, is an internal Rossby number. 

\citetalias{YAMAMOTOYODEN2013} define $m_0(z)$ so that \begin{equation}
    m_0(z)=\int^{\frac{\pi}{2}}_0 m_r(\vartheta,z)\cos\vartheta\,\text{d}\vartheta,
\end{equation}
where $m_r(\vartheta,z)=u(\vartheta,z)a\cos\vartheta$ is the relative component of specific axial angular momentum $m$, and assume $m_0(z)$ has the following vertical structure:\begin{equation}
    m_0(z) = Ua\left[1-\cos\left(\frac{\pi z}{H}\right)\right]/2,
\end{equation}
where $H$ is the height at the model top. 

As \citetalias{YAMAMOTOYODEN2013} specify a vertical structure for their model, we may relate $S^{\prime}$, which measures super-rotation at the model top, to the global super-rotation index $S$, \begin{equation}
    S=\frac{\frac{Ua}{2}\int^{H}_{0}\left[1-\cos\left(\frac{\pi z}{H}\right)\right]\,\text{d}z}{\varOmega a^{2}H\int^{\frac{\pi}{2}}_{0}\cos^{3}\vartheta\,\text{d}\vartheta} = \frac{3}{4}S^{\prime}.  
\end{equation}

\balance

\setlength{\bibsep}{2pt}
\begin{small}
    \bibliography{references}
    \bibliographystyle{ametsoc2014_var}
\end{small}

\end{document}